\documentclass[12pt]{article}
\usepackage{amsmath,amssymb,amsfonts,amsthm,amscd}
\usepackage[english]{babel}
\usepackage[toc]{appendix}
\usepackage{graphicx}
\usepackage{float}
\newcommand{\be}{\begin{equation}}
\newcommand{\ee}{\end{equation}}
\newcommand{\bea}{\begin{eqnarray}}
\newcommand{\eea}{\end{eqnarray}}
\newcommand{\nn}{\nonumber \\}
\newcommand{\p}[1]{(\ref{#1})}
\newcommand{\lb}{\label}
\newcommand{\sfrac}[2]{{\textstyle\frac{#1}{#2}}}

\begin{document}
\begin{titlepage}
\vspace*{0.7cm}

\begin{center}
\baselineskip=16pt {\Large\bf New realizations of the supergroup $D(2,1;\alpha)$ \\
\vspace{0.2cm}

in ${\cal N}=4$ superconformal mechanics }

\vskip 0.3cm {\large {\sl }} \vskip 10.mm {\bf $\;$  S. Fedoruk$^{\,\ddagger\,a}$,  $\;$ E. Ivanov$^{\,b}$
}
\vspace{1cm}

{\it Bogoliubov Laboratory of Theoretical Physics, JINR, \\
141980 Dubna, Moscow Region, Russia\\
}
\end{center}
\vfill

\par
\begin{center}
{\bf Abstract}
\end{center}
We present new explicit realizations of the most general ${\cal N}=4, d=1$ superconformal symmetry $D(2,1;\alpha)$
in the models of ${\cal N}=4$ superconformal mechanics based on the reducible multiplets $({\bf 1, 4, 3}) \oplus ({\bf 0, 4, 4})$,
$({\bf 3, 4, 1}) \oplus ({\bf 0, 4, 4})$   and   $({\bf 4, 4, 0}) \oplus ({\bf 0, 4, 4})$. We start from the manifestly supersymmetric superfield actions
for these systems and then descend to  the relevant off- and on-shell component actions from which we derive the $D(2,1;\alpha)$
(super)charges by the Noether procedure. Some peculiarities of these realizations of $D(2,1;\alpha)$ are discussed. We also construct
a new $D(2,1;\alpha)$ invariant system by joining the multiplets $({\bf 3, 4, 1})$ and $({\bf 4, 4, 0})$ in such a way that they
interact with each other through an extra $({\bf 0, 4, 4})$ multiplet.
New fermionic conformal couplings appear as the result of elimination of the appropriate auxiliary fields.

\vspace{2.5cm}
\noindent PACS: 03.65-w, 11.30.Pb, 12.60.Jv, 04.60.Ds\\
\smallskip
\noindent Keywords: supersymmetry, superfields, superconformal mechanics

\begin{quote}
\vfill \vfill \vfill \vfill \vfill \hrule width 5.cm \vskip 2.mm
{\small
\noindent $^{\ddagger}$ On leave of absence from V.N.\,Karazin Kharkov National University, Ukraine.\\
\noindent $^a$ fedoruk@theor.jinr.ru\\
\noindent $^b$ eivanov@theor.jinr.ru
}
\end{quote}
\end{titlepage}


\numberwithin{equation}{section}

\section{Introduction}
Superconformal mechanics (SCM) \cite{AP} - \cite{leva}  has plenty of applications \cite{nscm} - \cite{ISTconf}. For instance, the SCM models
can be identified with the denominator theories in the AdS$_2/$CFT$_1$ version of the general AdS$/$CFT correspondence and used for
the microscopic description of the extremal black holes \cite{nscm}-\cite{BMSV}, \cite{BKSS}, \cite{Galaj}.
Various versions of the ${\cal N}=4$ SCM are of special interest in these respects, since they provide an explicit description
of the massive  ${\cal N}=4$ superparticle
moving near the horizon of an extreme Reissner-Nordstr\"om black hole (see, e.g., \cite{nscm}, \cite{GTow}, \cite{BMSV}, \cite{Galaj}).
An important class of
the multiparticle SCM models is constituted by integrable superconformal Calogero-type systems \cite{GTow}, \cite{GLP2}, \cite{FIL}, \cite{KLP}, \cite{kl}.
A review of possible implications of SCM
in various domains, including ${\cal N}=4$ case, and additional references can be found  in \cite{superc}.

The most general ${\cal N}=4, d=1$ superconformal group is the exceptional supergroup $D(2,1;\alpha)$ \cite{FRS}, \cite{VP}.
At $\alpha =0, -1$ it reduces
to the semi-direct product $PSU(1,1|2) \rtimes SU(2)$ and at $\alpha = -\frac12$ to the supergroup $OSp(4|2)$\footnote{The isomorphic superalgebras
are related by the redefinition $\alpha \rightarrow -(1 + \alpha)\,$.}. The realizations  of  $D(2,1;\alpha)$ in the models
of supersymmetric mechanics were a subjects  of many works
(see, e.g., \cite{stro1}-\cite{BIKL}, \cite{BK1}-\cite{KuTo}, \cite{Govil}, \cite{PlWi}, \cite{ISTconf},
\cite{superc} and references therein)\footnote{For implications of $D(2,1;\alpha)$ in string theory and AdS$/$CFT
correspondence see, e.g., \cite{others}.}.
As a rule,  the realizations
on one or another fixed type of the irreducible ${\cal N}=4, d=1$ supermultiplet were considered. Recently,
the study of $SU(1,1|2)$ superconformal
systems including some pairs of such multiplets was initiated in Ref. \cite{Galaj}. Some interesting links with the ${\cal N}=2, d=5$
supergravity were established there. One of the basic points of the construction in \cite{Galaj} was the inclusion of couplings with the
fermionic ${\cal N}=4$ multiplets ${\bf (0,4,4)}$ which do not enlarge the dimension of the target bosonic manifold.

In the present paper we  study the analogous realizations of $D(2,1;\alpha)$  as distinct from \cite{Galaj},
where the particular  $SU(1,1|2)$ case was treated. Another new point of our consideration  is that in all cases we
start from the manifestly ${\cal N}=4$ supersymmetric off-shell superfield description of the relevant multiplet pairs
and write down the off-shell component Lagrangians, while in \cite{Galaj} only on-shell versions of the latter were addressed.
Keeping the relevant auxiliary fields in the combined actions of different pairs allows one to get more general on-shell component
actions after elimination of these fields by their algebraic equations of motion.\footnote{As distinct from the $D(2,1;\alpha)$ invariant systems with
the so-called ${\cal N}=4$ spin multiplets \cite{FIL2}, \cite{FIL3}, in our case
all bosonic fields of physical dimension are dynamical.}

Since the natural off-shell description of the multiplets considered in \cite{Galaj} and in the present paper is achieved
in the framework of ${\cal N}=4, d=1$ harmonic superspace \cite{IL}, we start in section 2 with recalling the basics of this approach.
Then, in section 3, we present the superfield and component descriptions of the fermionic multiplet ${\bf (0, 4, 4)}\,,$ which
is the common part of all systems considered in \cite{Galaj} and here. In section 4 we describe the $D(2,1;\alpha)$ invariant
system of interacting ${\bf (1, 4, 3)}$  and  ${\bf (0, 4, 4)}\,$ multiplets, both in the superfield and the component formulations,
and give the precise form of the $D(2,1;\alpha)$ generators for this case. In sections 5 and 6 we do the same for the multiplet pair
${\bf (3, 4, 1)}$  and  ${\bf (0, 4, 4)}\,$, as well as for the pair ${\bf (4, 4, 0)}$  and  ${\bf (0, 4, 4)}\,$. In section 7, as
an example of the power of the off-shell approach, we present a new superconformal system involving the triple of the multiplets
${\bf (1, 4, 3)}$, ${\bf (4, 4, 0)}$   and  ${\bf (0, 4, 4)}$.  The elimination of the auxiliary fields in the corresponding  action
yields new fermionic terms which are absent in the actions of the relevant isolated pairs. Section 8 contains conclusions and outlook.

\section{${\cal N}=4, d=1$ harmonic superspace}
The harmonic analytic ${\cal N}{=}4, d=1$ superspace \cite{IL,1,2} as the one-dimensional version of the general harmonic superspace \cite{HSS}
is defined as the following coordinate set
\be
(\zeta, u) = (t_A, \theta^+, \bar\theta^+, u^\pm_i)\,, \quad u^{+ i}u_i^- =1\,.
\ee
These coordinates are related to the standard ${\cal N}{=}4, d{=}1$ superspace (central basis) coordinates
$z = ( t, \theta_i, \bar\theta^i)$, $(\overline{\theta_i}) = \bar\theta^i$
as
\be
t_A = t +i (\theta^+\bar\theta^- + \theta^-\bar\theta^+), \quad \theta^\pm
= \theta^iu^\pm_i\,, \;
 \bar\theta^\pm = \bar\theta^iu^\pm_i\,.
\ee
The ${\cal N}{=}4$ covariant spinor derivatives and their harmonic projections
are defined by
\bea
&& D^i = \frac{\partial}{\partial \theta_i} - i\bar\theta^i \partial_t\,, \;\;
\bar D_i = \frac{\partial}{\partial \bar\theta^i} - i\theta_i \partial_t\,,
\;\; \overline{(D^i)}
= -\bar D_i\,, \;\;\{D^i, \bar D_k \} = -2i\,\delta^i_k\partial_t\,,
\label{defD2} \\
&& D^\pm = u^\pm_i D^i\,,\quad \bar D^\pm = u^\pm_i \bar D^i\,, \quad
 \; \{D^+, \bar D^- \} = - \{D^-, \bar D^+ \}
= -2i\,\partial_{t_A}\,. \label{defD1}
\eea
In the analytic basis $z_A = ( t_A, \theta^\pm, \bar\theta^\pm, u^{\pm i})$, the derivatives $D^{+}$ and $\bar D^+$ are short,
\be
D^+ = \frac{\partial}{\partial \theta{}^-}\,, \quad \bar D^+ =
-\frac{\partial}{\partial \bar\theta{}^-}\,.
\ee
The analyticity-preserving harmonic derivative $D^{++}$ and its conjugate
$D^{--}$ are given by
\bea
&& D^{++}=\partial^{++}+2i\theta^+\bar\theta^+\partial_{t_{A}}
+\theta^+\frac{\partial}{\partial\theta^-}
+ \bar\theta^+\frac{\partial}{\partial\bar\theta^-}\,, \nn
&& D^{--}=\partial^{--}+2i\theta^-\bar\theta^-\partial_{t_{A}}
+\theta^-\frac{\partial}{\partial\theta^+}
+  \bar\theta^-\frac{\partial}{\partial\bar\theta^+}\,, \quad
\partial^{\pm\pm} = u^{\pm}_i\frac{\partial}{\partial u^{\mp}_i}\,,
\eea
and become the pure partial derivatives $\partial^{\pm\pm}$ in the central basis.
They satisfy the relations
\be
[D^{++},D^{--}]= D^{0}\,, \quad [D^0, D^{\pm\pm}] =
\pm 2 D^{\pm\pm}\,, \lb{DharmAl}\\
\ee
where $D^0$ is the operator counting external harmonic $U(1)$ charges.
The integration measures in the full harmonic superspace and
its analytic subspace are defined as
\begin{eqnarray}
&& \mu_H = dudtd^4\theta=dudt_{A}(D^-\bar D^-)(D^+\bar D^+)
=\mu_{A}^{(-2)}(D^+\bar D^+)\,,\nn
&& \mu_{A}^{(-2)}=dud\zeta^{(-2)}
=dudt_{A}d\theta^+d\bar\theta^+=dudt_{A}(D^-\bar D^-)\,. \label{measures}
\end{eqnarray}

The analytic subspace $(\zeta, u)$ is closed under the action of the
most general ${\cal N}=4, d=1$ superconformal group $D(2,1;\alpha)$
and its degenerate $D(2,1;\alpha=0)$ and $D(2,1;\alpha=-1)$ cases
which are reduced to the semi-direct products $PSU(1,1|2)\rtimes
SU(2)_{\rm ext}$. In what follows, we will need the  transformation
properties of some relevant quantities under the ``Poincar\'e'' and
conformal supersymmetry. The invariance under these transformations
is sufficient for ensuring the complete $D(2,1; \alpha)$ invariance
since the rest of the $D(2,1; \alpha)$ transformations is contained
in the closure of the conformal and manifest Poincar\'e ${\cal
N}{=}4, d{=}1$ supersymmetries.

In the ${\cal N}{=}4$ superfield approach, the invariance under the ordinary $d=1$ supertranslations
($\bar\varepsilon^i=\overline{(\varepsilon_i)}$)
\be
\delta t =i(\varepsilon_k\bar\theta^k - \theta_k\bar\varepsilon^k),\qquad
\delta
\theta_k=\varepsilon_k, \qquad \delta \bar\theta^k=\bar\varepsilon^k \lb{simple}
\ee
and
\be
\delta t_A =2i(\varepsilon^-\bar\theta^+ - \bar\varepsilon^-\theta^+),\qquad
\delta
\theta^+=\varepsilon^+, \qquad \delta \bar\theta^+=\bar\varepsilon^+\,, \lb{simple1}
\ee
where $\varepsilon^{\pm} = \varepsilon^iu^\pm_i\,, \;\bar{\varepsilon}^{\pm} =
\bar\varepsilon^iu^\pm_i\,$, is automatic.

The coordinate realization of the superconformal $D(2,1;\alpha)$ boosts is
as follows :
\begin{equation}  \label{sc-coor-c-b}
\delta^\prime t=i t(\theta_k \bar\eta^k + \bar\theta^k\eta_k) - (1+\alpha)\,\theta_i
\bar\theta^i (\theta_k \bar\eta^k + \bar\theta^k\eta_k)\,,
\end{equation}
\begin{equation}  \label{sc-coor-c-f1}
\delta^\prime \theta_i= -\eta_i t -2i\alpha \,\theta_i(\theta_k \bar\eta^k) +
2i(1+\alpha)\,\theta_i (\bar\theta^k\eta_k)- i(1+2\alpha)\,\eta_i (\theta_k
\bar\theta^k)\,,
\end{equation}
\begin{equation}  \label{sc-coor-c-f2}
\delta^\prime \bar\theta^i= -\bar\eta^i t -2i\alpha \,\bar\theta^i(\bar\theta^k\eta_k) +
2i(1+\alpha)\,\bar\theta^i (\theta_k \bar\eta^k)+ i(1+2\alpha)\,\bar\eta^i (\theta_k
\bar\theta^k)\,,
\end{equation}
\begin{equation}  \label{sc-coor-bo}
\delta^\prime t_A=\alpha^{-1}\Lambda_{sc} t_A\,,\qquad \delta^\prime u^+_i= \Lambda^{++}u^-_i\,,
\end{equation}
\begin{equation}  \label{sc-coor-fer}
\delta^\prime \theta^+= -\eta^+ t_A +2i(1+\alpha)\eta^- \theta^+\bar\theta^+\,,\qquad
\delta^\prime \bar\theta^+= -\bar\eta^+ t_A +2i(1+\alpha)\bar\eta^- \theta^+\bar\theta^+ \,,
\end{equation}
\begin{equation}  \label{sc-me1}
\delta^\prime (dtd^4\theta)=- (dtd^4\theta)\,\Lambda_0\,,\qquad \delta^\prime
\mu_H= \mu_H\left(2\Lambda_{sc} - (1+\alpha)\Lambda_0\right)\,,\qquad \delta^\prime
\mu^{(-2)}_A= 0\,,
\end{equation}
\begin{equation}  \label{sc-me2}
\delta^\prime\, \mu_A^{(-2)} = 0\,, \qquad
\delta^\prime\,du = du\,D^{--}\Lambda^{++}\,,
\end{equation}
\begin{equation} \label{DDConf}
\delta^\prime\, D^{++} = -\Lambda^{++} \, D^0\,,
\qquad \delta^\prime\, D^0 = 0\,.
\end{equation}

Here $\eta^{\pm} = \eta^iu^\pm_i\,, \;\bar\eta^{\pm} =
\bar\eta^iu^\pm_i\,$, $\bar\eta^i=\overline{(\eta_i)}\,,$ and
\begin{equation}  \label{def-La1}
\Lambda_{sc} = 2i\alpha (\bar\eta^-\theta^+ -\eta^-\bar\theta^+ )\,,\quad
\Lambda^{++} = D^{++}\Lambda_{sc}=2i\alpha (\bar\eta^+\theta^+ - \eta^+\bar\theta^+ )\,, \quad
D^{++}\Lambda^{++}=0\,,
\end{equation}
\begin{equation}  \label{def-La2}
\Lambda_0 = \alpha^{-1}\left(2\Lambda_{sc} - D^{--} \Lambda^{++}\right) = 2i (\theta_k \bar\eta^k +
\bar\theta^k\eta_k)\,, \qquad D^{++}\Lambda_0=0\,.
\end{equation}
The symbol $\sim$ means the generalized tilde-conjugation \cite{HSS}. With such definitions, all the coordinate
transformations contain no singularities in the degenerate $\alpha = 0$ or $\alpha =-1$ cases.

\section{The fermionic multiplet $({\bf 0, 4, 4})$}
This multiplet is the fermionic analog of the multiplet $({\bf 4, 4, 0})$. It is described off shell by the fermionic
analytic superfield $\Psi^{+ A}\,, \; A=1,2\,$,
$\widetilde{(\Psi^{+ A})} = \Psi_A^+\,$, satisfying the constraint \cite{IL}:
\be
D^{++}\Psi^{+ A} = 0\qquad \Rightarrow \qquad \Psi^{+ A} =
\phi^{iA}u^+_i + \theta^+ F^A + \bar\theta^+ \bar{F}^A
- 2i\theta^+\bar\theta^+ \dot{\phi}{}^{i A}u^-_i\,,  \label{PsiConstr}
\ee
where $(\overline{\phi^{iA}})=-\phi_{iA}$, $(\overline{F^A})=\bar{F}_A$. On the index $A$, the appropriate $SU(2)_{PG}$ group acts.
Its generators  commute with the ${\cal N}=4$ supersymmetry and $D(2,1;\alpha)$ generators.
The requirement of superconformal
covariance of the constraint \p{PsiConstr} uniquely fixes the superconformal
$D(2,1;\alpha)$ transformation rule of $\Psi^{+ A}$, for any $\alpha$, as
\be
\delta_{sc}\,\Psi^{+ A} = \Lambda_{sc}\, \Psi^{+ A}\,.\label{traNPsi1}
\ee

Off-shell transformations of component fields are the following
\be\label{tr-comp-044}
\begin{array}{c}
\delta\phi^{iA}=-\left( \omega^i F^A+\bar\omega^i \bar F^A\right), \\ [6pt]
\delta F^A=2i\,\bar\omega^k\dot\phi^A_k +2i\alpha\,\bar\eta^k\phi^A_k\,,\qquad
\delta \bar F_A=2i\,\omega_k\dot{\bar\phi}^{k}_A +2i\alpha\,\eta_k {\bar\phi}^{k}_A\,,
\end{array}
\ee
where
$$
\omega_i= \varepsilon_i -t\,\eta_i\,,\qquad
\bar\omega^i= \bar\varepsilon^i - t\,\bar\eta^i\,.
$$

In the central basis, the constraint \p{PsiConstr} and the analyticity conditions $D^+\Psi^{+ A}{=} \bar D^+\Psi^{+ A}{=}\,0$
imply
\bea
\Psi^{+ A}(z, u) = \Psi^{i A}(z)u^+_i\,,  \label{CBpsi}\\ [7pt]
D^{(i} \Psi^{k) A}(z) = \bar D^{(i} \Psi^{k) A}(z) = 0\,.\label{ConstrPsi2}
\eea

The free action of $\Psi^{+ A}$,
\be
S^{(\Psi)}_{free} = \sfrac12 \int du d\zeta^{(-2)}\,\Psi^{+ A}\Psi^{+}_{A} =  \int dt \left(-i\phi^{iA}\dot\phi_{iA} + F^A\bar{F}_A \right)\,,
\label{Freepsi}
\ee
is not invariant under $D(2,1;\alpha)$, except for the special case of
$\alpha{=}0\,$, in which we will not be too interested. As we will see, the superconformal versions of the free $\Psi^{+ A}$ action,
which are valid for any $\alpha$, can be constructed by coupling this multiplet to those considered in the next sections.

The only additional ${\cal N}{=}4$ invariant is the appropriate Fayet-Iliopoulos (FI)-type term
\be
S^{(\Psi)}_{FI} = \gamma\int du d\zeta^{(-2)} \left( \theta^+ \xi_A\Psi^{+ A}
+ \bar\theta^+ \bar\xi^A\Psi^+_A \right)
= \gamma\int dt \left( \bar\xi_A F^A-\xi^A \bar F_A  \right), \label{FIpsi}
\ee
$\xi_A$, $\bar\xi^A=(\overline{\xi_A})$ being $SU(2)_{PG}$ breaking constants.
It is superconformal at $\alpha{=}-1$ \cite{2}.

\section{The multiplet pair $({\bf 1, 4, 3})\oplus ({\bf 0, 4, 4})$}

\subsection{The multiplet $({\bf 1, 4, 3})$}
The off-shell multiplet ${\bf (1, 4, 3)}$ is described by
a real ${\cal N}{=}4$  superfield $X(z)$ obeying the constraints \cite{leva}
\be
D^iD_i X = \bar D_i\bar D^i X = 0\,, \quad [D^i, \bar D_i] X = 0\,. \label{Uconstr1}
\ee
Solution of these constraints is provided by
\begin{equation}  \label{sing-X0-WZ}
X(t,\theta_i,\bar\theta^i)= r + \theta_i\varphi^i +
\bar\varphi_i\bar\theta^i +
i\theta^i\bar\theta^k A_{ik}-{\textstyle\frac{i}{2}}(\theta)^2\dot{\varphi}_i\bar\theta^i
-{\textstyle\frac{i}{2}}(\bar\theta)^2\theta_i\dot{\bar\varphi}{}^i +
{\textstyle\frac{1}{4}}(\theta)^2(\bar\theta)^2 \ddot{r}\,,
\end{equation}
where $(\overline{r})=r$, $(\overline{\varphi^i})=\bar\varphi_i$, $(\overline{A^{ik}})=A_{ik}=A_{(ik)}$ and
$(\theta)^2=\theta_k \theta^k$, $(\bar\theta)^2=\bar\theta^k \bar\theta_k$.

The same constraints \p{Uconstr1} rewritten in harmonic superspace read
\be
D^{++}X =0\,, \;\; D^+D^-X = \bar D^+\bar D^-X = 0\,, \quad
\left(D^+\bar D^- + \bar D^+D^-\right)X = 0\,.\label{Uconstr2}
\ee
The extra harmonic constraint guarantees the harmonic independence of
$X$ in the central basis.

As was shown in \cite{2}, this multiplet can be also described in terms
of the real analytic gauge superfield ${\cal V}(\zeta, u)$ with
the abelian gauge transformation
\be
{\cal V} \;\;\Rightarrow \;\; {\cal V}{\,}' = {\cal V} + D^{++}\Lambda^{--}\,, \quad
\Lambda^{--} = \Lambda^{--}(\zeta, u)\,.\label{VgaugeT}
\ee
In the Wess-Zumino (WZ) gauge just the irreducible ${\bf (1, 4, 3)}$ content remains
\be
{\cal V}_{WZ}(\zeta, u) = r(t_A) -2 \theta^+\varphi^i(t_A) u^-_i
-2 \bar\theta^+ \bar\varphi^i(t_A) u^-_i +
3i \theta^+\bar\theta^+ A^{(ik)}(t_A)u^-_iu^-_k\,. \label{WZV1}
\ee
No residual gauge freedom is left. The original superfield $X(z)$ is
related to ${\cal V}(\zeta, u)$ by
\be
X(t, \theta^i, \bar\theta_k)=
\int du\, {\cal V}\left(t +2i\theta^i\bar\theta^k u^+_{(i}u^-_{k)}\,,\, \theta^iu^+_i,
\bar\theta^ku^+_k\,,\, u^\pm_l\right). \label{DefU2}
\ee
The constraints \p{Uconstr1} are recovered as a consequence of the harmonic
analyticity of  ${\cal V}\,$,
\be
D^+{\cal V} = \bar D^+{\cal V} = 0\,.\label{Vanalit}
\ee

The transformation properties of the superfields $X$ and ${\cal V}$ under
the $D(2,1; \alpha)$ conformal supersymmetry are defined by
\bea
\delta_{sc}\, X = -\alpha\,\Lambda_0\,X\,, \qquad \delta_{sc}\,{\cal V} =
-2\Lambda_{sc}\,{\cal V}\,.\label{vVconf}
\eea
The full set of the component fermionic transformations obtained from \p{simple} - \p{sc-me1} reads
\begin{equation}\label{tr-comp-143}
\begin{array}{c}
\delta r=-\omega_i\varphi^i+ \bar\omega^i\bar\varphi_i\,, \\ [6pt]
\delta \varphi^i=i \bar\omega^i\dot r-i \bar\omega_k A^{ki}-2i\alpha\,\bar\eta^i r  \,,
\qquad
\delta \bar\varphi_i=-i \omega_i\dot r -i \omega^k A_{ki} +2i\alpha\,\eta_i r
\,,\\ [6pt]
\delta A_{ik}=-2\left( \omega_{(i}\dot\varphi_{k)} +\bar\omega_{(i} \dot{\bar\varphi}_{k)}\right)
+2(1+2\alpha)\left( \eta_{(i}\varphi_{k)} +\bar\eta_{(i} {\bar\varphi}_{k)}\right),
\end{array}
\end{equation}
where $\omega_i = \varepsilon_i - t\, \eta_i\, $ and $\bar\omega^i =
\bar\varepsilon^i - t\, \bar\eta^i\, $.

The general $\varepsilon_i, \bar\varepsilon_k$-invariant superfield action of the multiplet ${\bf (1, 4, 3)}$ is written as
\be
S_{gen}^{(X)} = \int dtd^4\theta\, {\cal L}_{gen}(X)\,.\label{genact_v}
\ee
The action invariant under $D(2,1; \alpha)$ (except for the special value of $\alpha{=}0$)
is \cite{ikl1}
\be \label{conf_v}
S_{sc}^{(X)} =  -{\textstyle\frac{1}{8(1+\alpha)}} \int dtd^4\theta\left(X^{-1/\alpha}-X\right).
\ee
For the correct $d{=}1$ field theory interpretation, one must assume that $X$
develops a non-zero background value, $X = 1 + \ldots \,$. Note that
${\int} dtd^4\theta \, X$ is an integral of total derivative and does not contribute for $\alpha \neq -1\,$.
We add this term for ensuring the correct limit $\alpha{=}-1$, in which \p{conf_v} is reduced to
\bea
S^{(X)(\alpha =-1)}_{sc} = -{\textstyle\frac{1}{8}}\int dt d^4\theta\, X\,\log X\,.
\eea

Using the coordinate and superfield transformations \p{sc-coor-c-b} - \p{sc-me2}, \p{vVconf},
it is easy to check the $D(2,1; \alpha)$ invariance of the action \p{conf_v} and the covariance of the relation \p{DefU2}.

In the component field notation  the action \p{conf_v} takes the form
\begin{eqnarray}
S^{(X)}_{sc} &=& \textstyle{\frac{1}{8\alpha^2}}\,\displaystyle{\int} dt\,
r^{-\frac{1}{\alpha}-2}\,\left[ \dot r\dot r +i \left(\bar\varphi_k \dot\varphi^k
-\dot{\bar\varphi}_k \varphi^k \right) +{\textstyle\frac{1}{2}}A^{ik}A_{ik}\, \right] \nn
&&  -\, \textstyle{\frac{i}{8\alpha^2}}\, (\textstyle{\frac{1}{\alpha}}+2)
\,\displaystyle{\int} dt\, r^{-\frac{1}{\alpha}-3}\, A^{ik} \varphi_{(i}\bar\varphi_{k)}  \nn
&& - \,
\textstyle{\frac{1}{24\alpha^2}}\,
(\textstyle{\frac{1}{\alpha}}+2)(\textstyle{\frac{1}{\alpha}}+3)\, \displaystyle{\int} dt\,
r^{-\frac{1}{\alpha}-4}\, \varphi^{i}\bar\varphi^{k} \varphi_{(i}\bar\varphi_{k)} \,.\label{4N-X-WZ}
\end{eqnarray}

One can also construct an ${\cal N}{=}4$ supersymmetric FI term
\be
S_{FI}^{(X)} = i\int du d\zeta^{(-2)}\, c^{+2}\,
{\cal V}\,, \qquad c^{+2} = c^{ik}u^+_iu^+_k\,, \qquad [c] = cm^{-1}.\label{FI1}
\ee
It generates a scalar potential after elimination of the auxiliary field
$A^{(ik)}$ in the sum of
\p{genact_v} and \p{FI1}. The FI term is superconformal only for the special choice
$\alpha{=}0$ \cite{2}. At this value of $\alpha$, the corresponding scalar potential
is the standard inverse-square conformal potential $\sim c^2 = c^{ik}c_{ik}$. The $\alpha=0$
conformal kinetic sigma-model term can be constructed at cost of modifying the superconformal transformation law of $X$ for this value
of $\alpha$ as
\bea
\delta_{sc}^{(\alpha=0)} X = -\Lambda_0\,, \quad
\delta_{sc}^{(\alpha=0)} {\cal V} = 4i\left(\bar\varepsilon^-\theta^+ - \varepsilon^-\bar\theta^+\right) \label{modTrans}, \lb{mod0}
\eea
under which the constraints \p{Uconstr1} are still covariant. Then the superconformal kinetic term
is given by
\bea
S^{(X)(\alpha =0)}_{sc} = \int dt d^4\theta\, e^X\,.\lb{Actalphazero}
\eea
The FI term \p{FI1} remains invariant under the modified transformation rule \p{mod0}.

For the choice of  $\alpha =-1$, the conformal component potential can be secured by modifying the last constraint in \p{Uconstr1} as
\bea
[D^i, \bar D_i] X = 0\; \Rightarrow \; [D^i, \bar D_i] X = c\,,
\eea
where $c$ is a real constant. At $\alpha =-1$ the modified constraints are still superconformally covariant. The conformal potential
appears with the strength $\sim c^2\,$.  Note that the superfield $X$ subjected to the modified constraints is related to $X_{(c=0)}$ as
\bea
X = X_{(c=0)} + \frac{1}{2}\,c\,\bar\theta^i\theta_i =  X_{(c=0)} +
\frac{1}{2}\,c\left(\bar\theta^+\theta^- - \bar\theta^-\theta^+ \right). \label{tildeU}
\eea
For $v_{(c=0)}$ there is still valid the prepotential representation \p{DefU2}. For $X$ to have the correct $\alpha=-1$ transformation law,
$\delta_{sc}^{(\alpha =-1)} X = \Lambda_0\,X$, the transformation rule of the prepotential ${\cal V}$  under the Poincar\'e and conformal supersymmetries
should be modified as
\bea
\delta_{sc}^{(\alpha =-1)}{\cal V} = -2\Lambda_{sc}^{(\alpha =-1)}{\cal V}+ c\left[(\epsilon^- +t_A\varepsilon^-)\bar\theta^+ + (\bar\epsilon{}^-
+t_A\bar\varepsilon^-)\theta^+ \right]. \label{modTransI}
\eea

\subsection{Superconformal coupling of multiplets $({\bf 1, 4, 3})$ and  $({\bf 0, 4, 4})$}

Using the  description of the multiplet $({\bf 1, 4, 3})$ through the analytic prepotential ${\cal V}$,  it is
easy to construct its superconformal coupling to $\Psi^{+ A}$ \cite{2}
\be
S^{(X,\Psi)}_{sc} =\sfrac12\,b\int du d\zeta^{(-2)}\,{\cal V}\, \Psi^{+ A}\Psi^{+}_{ A}\,.
\label{Confpsi}
\ee
This action is superconformal at any $\alpha \neq 0\,$\footnote{It is not invariant under the modified $\alpha=0$
transformation of ${\cal V}$, eq. \p{mod0}, as well as under \p{modTransI} corresponding to the $c\neq 0, \alpha =-1$
version of $({\bf 1, 4, 3})$ multiplet. So these options are excluded.} and it also respects
the gauge invariance
\p{VgaugeT} as a consequence of the constraint \p{PsiConstr}. Assuming that
${\cal V} = 1 + \widetilde{{\cal V}}$, $v = 1 + \widetilde{v}$, \p{Confpsi} can be
treated as a superconformal generalization of the free action \p{Freepsi}.
An analysis based on dimensionality and on the Grassmann character of the superfields
$\Psi^{+ A}, \Psi^{- A} = D^{--}\Psi^{+ A}$ shows that the coupling
\p{Confpsi} is the only possible coupling of this fermionic multiplet to the
multiplet ${\bf (1,4,3)}$
preserving the canonical number of time derivatives
in the component action
(no more than two for bosons and no more than one for fermions).

It is easy to obtain the component-field representation of \p{Confpsi}
\be \label{coup-143-comp}
S^{(X,\Psi)}_{sc} = b \int dt\left[r\left(-i\phi^{iA}\dot{\phi}_{iA} + F^A\bar F_A\right) +
\sfrac{i}{2}A^{ik}\,\phi^A_i\phi_{kA}  +\bar\varphi{}^k\phi_{kA}\, F^A
- \varphi_k\phi^{kA}\,\bar F_A  \right] .
\ee
After summing up this action with the component superconformal ${\bf (1,4,3)}$ action \p{4N-X-WZ} and eliminating the auxiliary fields $A_{ik}$
and $F^A, \bar{F}_A\,$,
\be
\begin{array}{c}
A_{ik}= i\left({\textstyle\frac{1}{\alpha}+2} \right)r^{-1}\varphi_{(i}\bar\varphi_{k)} -
4b\alpha^2i r^{\frac{1}{\alpha}+2}\phi_{(i}^{A}\phi_{k)A}\,,
\\ [6pt]
F^A= r^{-1}\varphi_{k}\phi^{kA}\,,\qquad \bar{F}_A =-r^{-1}\bar\varphi^{k}\phi_{kA}\,,
\end{array}
\ee
we obtain the total on-shell superconformal action as
\begin{eqnarray}
S_{(X+\Psi)} &=&
\textstyle{\frac{1}{8\alpha^2}}\,\displaystyle{\int} dt\,
r^{-\frac{1}{\alpha}-2}\,\left[ \dot r\dot r +i \left(\bar\varphi_k
\dot\varphi^k -\dot{\bar\varphi}_k \varphi^k \right) \right] -i b
\int dt\,r\,\phi^{iA}\dot{\phi}_{iA} \nonumber
\\ &&    + \,
\displaystyle{\int} dt\Big[
\textstyle{\frac{1+2\alpha}{48\,\alpha^4}}\,
r^{-\frac{1}{\alpha}-4}\, \varphi_{(i}\bar\varphi_{k)}
\varphi^{i}\bar\varphi^{k}  -
\textstyle{\frac{b}{2\alpha}}\,r^{-1}\varphi_{(i}\bar\varphi_{k)}\phi^{iA}\phi^{k}_A
\nonumber
\\ &&\hspace{1.5cm} +\, \alpha^2 b^2 r^{\frac{1}{\alpha}+2}\,\phi_{(i}^{A}\phi_{k)A}\phi^{iB}\phi^{k}_B\Big].\label{totalvpsi}
\end{eqnarray}

Redefining the field variables as
\begin{equation}\label{4N-nZ}
x = r^{-\frac{1}{2\alpha}} \,, \qquad \psi_{k} =
-{\textstyle\frac{1}{2\alpha}}\,r^{-\frac{1}{2\alpha}-1} \varphi_{k}\,, \qquad
\chi^{iA} = \sqrt{2br}\,\phi^{iA}\,,
\end{equation}
we cast the action \p{totalvpsi} into the convenient form
\begin{eqnarray}
S_{(X+\Psi)} &=& \textstyle{\frac{1}{2}}\,\displaystyle{\int} dt\,
\left[ \dot x\dot x +i \left(\bar\psi_k \dot\psi^k -\dot{\bar\psi}_k
\psi^k  \right) -i\chi^{iA}\dot{\chi}_{iA}\right] \nonumber
\\
&&    + \, \displaystyle{\int} dt\,\frac{1}{x^{2}} \Big[
\textstyle{\frac{1}{3}}\,(1+2\alpha) \psi_{(i}\bar\psi_{k)}
\psi^{i}\bar\psi^{k}  -
\alpha\,\psi_{(i}\bar\psi_{k)}\chi^{iA}\chi^{k}_A \nonumber
\\
&& \hspace{2cm} +\, \textstyle{\frac{1}{4}}\,\alpha^2
\,\chi_{(i}^{A}\chi_{k)A}\chi^{iB}\chi^{k}_B\Big].
\label{totalvpsi-1}
\end{eqnarray}
It is invariant under the following on-shell supersymmetry
transformations
\begin{equation}\label{2str-X}
\delta x=-\omega_i\psi^i+ \bar\omega^i\bar\psi_i\,,
\end{equation}
\begin{equation}\label{2str-Psi1}
\delta \psi^i=i \bar\omega^i\dot x + i\bar\eta^i x
-(1+2\alpha)\,\frac{\omega_k \psi^{k}\psi^{i}+\bar\omega_k \psi^{k} \bar\psi^{i} }{x}
+\alpha\,\frac{\bar\omega_k \chi^{A(i} \chi^{k)}_A}{x} \,,
\end{equation}
\begin{equation}\label{2str-Psib}
\delta \bar\psi_i=-i \omega_i\dot x-i\eta_i x
+(1+2\alpha)\,\frac{\bar\omega^k \bar\psi_{k}\bar\psi_{i}+\omega^k \bar\psi_{k} \psi_{i} }{x}
+\alpha\,\frac{\omega^k \chi^{A}_{(i} \chi_{k)A}}{x}\,,
\end{equation}
\begin{equation}\label{2str-Z}
\delta \chi^{iA}=-2\alpha\,\frac{\omega^{(i}\psi^{k)} +\bar\omega^{(i}\bar\psi^{k)}}{x}\, \chi^{A}_k\,.
\end{equation}
{}From \p{2str-Z} we observe that the supersymmetry transformation of $\chi^{iA}$ looks as a field-dependent
$SU(2)$ transformation. At $\alpha = -1$, the action \p{totalvpsi-1} is reduced to the corresponding on-shell
action from Ref. \cite{Galaj}.

Using the component $D(2,1;\alpha)$ transformations \p{2str-X}, \p{2str-Psi1}, \p{2str-Psib} and \p{2str-Z},
one can find the related Noether (super)charges
\begin{equation}\label{Q-cl}
{Q}^i =p\, \psi^i- i x^{-1}\psi_{k}\,\Big[{\textstyle\frac{2}{3}}\,(1+2\alpha)\,\psi^{(i}\bar\psi^{k)}
- \alpha\,\chi^{iA} \chi^{k}_A \Big]\, ,
\end{equation}
\begin{equation}\label{Qb-cl}
\bar{Q}_i=p\, \bar\psi_i+i x^{-1}\bar\psi^{k}\,\Big[{\textstyle\frac{2}{3}}\,(1+2\alpha)\,\psi_{(i}\bar\psi_{k)}
- \alpha\,\chi_i^{A} \chi_{kA} \Big]
\end{equation}
and
\begin{equation}\label{S-cl}
{S}^i =x \psi^i - t\,{Q}^i,\qquad \bar{S}_i=x\bar\psi_i-t\,\bar{Q}_i\,,
\end{equation}
where $ p\equiv \dot x$, and then check that they generate the classical $D(2,1;\alpha)$ superalgebra.

To this end, we define the non-vanishing canonical Dirac brackets (at equal times) as
\begin{equation}\label{CDB}
\{x, p\}_{{}_D}= 1, \qquad \{\psi^i, \bar\psi_j\}_{{}_D}=
-i\,\delta^i_j\,,\qquad
\{\chi^{iA}, \chi^{jB}\}_{{}_D}= i\,\epsilon^{ij}\epsilon^{AB},
\end{equation}
where we adopted the convention $\epsilon_{12} = \epsilon^{21} = 1$. Using (\ref{CDB}),  we arrive at the following Dirac brackets
for  fermionic generators
\begin{equation} \label{DB-QQ}
\begin{array}{c}
\{{Q}^{i}, \bar{Q}_{k}\}_{{}_D}= -2i\delta^{i}_{k}H\,,\qquad  \{{Q}^{i}, {Q}^{k}\}_{{}_D}=0
\,,\qquad  \{\bar{Q}_{i}, \bar{Q}_{k}\}_{{}_D}=0\,, \\ [7pt]
\{{S}^{i}, \bar{S}_{k}\}_{{}_D}= -2i\delta^{i}_{k}K\,,\qquad  \{{S}^{i}, {S}^{k}\}_{{}_D}=0
\,,\qquad  \{\bar{S}_{i}, \bar{S}_{k}\}_{{}_D}=0\,,
\\ [7pt]
\{{Q}^{i}, {S}^{k}\}_{{}_D}= 2i(1+\alpha)\,\epsilon^{ik}\,I\,,
\qquad
\{\bar{Q}_{i}, \bar{S}_{k}\}_{{}_D}=-2i(1+\alpha)\,\epsilon_{ik}\,\bar I\,,
\\ [7pt]
\{{Q}^{i}, \bar{S}_{k}\}_{{}_D}= 2i\delta^{i}_{k}\,D+2i\alpha\, J^i{}_k-2i(1+\alpha)\,\delta^{i}_{k}\,I_3\,,
\\ [7pt]
\{\bar{Q}_{i}, {S}^{k}\}_{{}_D}=2i\delta_{i}^{k}\,D-2i\alpha\, J_i{}^k+2i(1+\alpha)\,\delta_{i}^{k}\,I_3\,.
\end{array}
\end{equation}
Here, the bosonic generators
\begin{eqnarray}\label{H-cl}
H &=&{\textstyle\frac{1}{2}}\,p^2  + x^{-2}\Big[{\textstyle\frac{1}{4}}\,(1+2\alpha) \,
\psi_{i}\psi^{i}\,\bar\psi_{k} \bar\psi^{k}
+\alpha  \,\psi_{i}\bar\psi_{k}\chi^{iA}\chi^{k}_A
-{\textstyle\frac{1}{4}}\,\alpha^2\,\chi_{i}^{A}\chi^{}_{kA}\chi^{iB}\chi^{k}_B \Big]\,,
\\ \label{K-cl}
K &=&{\textstyle\frac{1}{2}}\,x^2  -t\,x p +
    t^2\, H\,,
\\ \label{D-cl}
D &=&-{\textstyle\frac{1}{2}}\,x p + t\, H\,,
\end{eqnarray}
\begin{equation} \label{I-cl}
I = {\textstyle\frac{i}{2}}\, \psi_k\psi^k\,,\qquad
\bar I = {\textstyle\frac{i}{2}}\, \bar\psi_k\bar\psi^k\,,\qquad
I_3 ={\textstyle\frac{i}{2}}\,  \psi_k\bar\psi^k\,,
\end{equation}
\begin{equation} \label{T-cl}
J^{ik} = -i\left[ \psi^{(i}\bar\psi{}^{k)}-{\textstyle\frac12}\,\chi^{iA}\chi^k_{A}\right]
\end{equation}
form the algebra
\begin{equation}\label{DB-T-1}
\{H, K\}_{{}_D}= 2D\,,\qquad
\{H, D\}_{{}_D}= H\,,\qquad
\{K, D\}_{{}_D}= -K\,,
\end{equation}
\begin{equation}\label{DB-I}
\{I, \bar I\}_{{}_D}= 2I_3\,,\qquad
\{I, I_3\}_{{}_D}= I \,,\qquad
\{\bar I, I_3\}_{{}_D}= -\bar I\,,\qquad
\end{equation}
\begin{equation}\label{DB-J}
\{J^{ij}, J^{kl}\}_{{}_D}=- \epsilon^{ik}J^{jl} -\epsilon^{jl}J^{ik}\,.
\end{equation}

Next, introducing the quantities
\begin{equation}\label{not-Q}
{Q}^{11^\prime i}={S}^{i}\,,\qquad {Q}^{12^\prime i}=\bar{S}^{i}\,,
\qquad\qquad
{Q}^{21^\prime i}={Q}^{i}\,,\qquad {Q}^{22^\prime i}=\bar{Q}^{i}\,,
\end{equation}
\begin{equation}\label{not-T}
T^{11}=K\,,\qquad T^{22}=H\,,\qquad T^{12}=-D\,,
\end{equation}
\begin{equation} \label{I-cl1}
I^{1^\prime 1^\prime} = I\,,\qquad
I^{2^\prime 2^\prime} = \bar I\,,\qquad
I^{1^\prime 2^\prime} =I_3 \,,
\end{equation}
we obtain that the closed superalgebra of the full set of generators takes the form
\begin{equation} \label{DB-Q-g}
\{{Q}^{ai^\prime i}, {Q}^{bk^\prime k}\}_{{}_D}= -2i\Big(\epsilon^{ik}\epsilon^{i^\prime
k^\prime} T^{ab}+\alpha \epsilon^{ab}\epsilon^{i^\prime k^\prime} J^{ik}-(1+\alpha)
\epsilon^{ab}\epsilon^{ik} I^{i^\prime k^\prime}\Big)\,,
\end{equation}
\begin{equation}\label{DB-T}
\{T^{ab}, T^{cd}\}_{{}_D}= -\epsilon^{ac}T^{bd} -\epsilon^{bd}T^{ac}\,,
\end{equation}
\begin{equation}\label{DB-I-c}
\{I^{i^\prime j^\prime}, I^{k^\prime l^\prime}\}_{{}_D}= -\epsilon^{i^\prime k^\prime}I^{j^\prime l^\prime}
-\epsilon^{j^\prime l^\prime}I^{i^\prime k^\prime}\,,
\end{equation}
\begin{equation}\label{DB-JQ}
\begin{array}{rcl}
\{T^{ab}, {Q}^{ci^\prime i}\}_{{}_D}&=&\epsilon^{c(a}{Q}^{b)i^\prime i} \,, \\ [7pt]
\{J^{ij},{Q}^{ai^\prime k}\}_{{}_D}&=&\epsilon^{k(i}{Q}^{ai^\prime j)}\,, \\ [7pt]
\{J^{i^\prime j^\prime},{Q}^{ak^\prime i}\}_{{}_D}&=&\epsilon^{k^\prime (i^\prime}{Q}^{aj^\prime ) i}\,.
\end{array}
\end{equation}
This is just the standard form of the superalgebra $D(2,1;\alpha)$.

Note that the relations \p{totalvpsi-1} - \p{DB-JQ} are also valid at $\alpha = 0$, in particular, the action \p{totalvpsi-1}
is invariant under the $\alpha=0$
version of the transformations \p{2str-X} - \p{2str-Z}. It looks somewhat paradoxical, since we started from the action \p{conf_v}
which is singular at $\alpha =0$.
The explanation is that it is just the free ${\bf (0, 4, 4)}$ action \p{Freepsi} which is superconformal at $\alpha =0$,
and so the superconformal action
of the system ${\bf (1, 4, 3)}\oplus {\bf (0, 4, 4)}$ at this value of $\alpha$ is the sum of the actions \p{Actalphazero}
and \p{Freepsi}, containing no interaction between
the two multiplets at all. In components, after some field redefinitions, this sum of actions is reduced on shell to the $\alpha =0$ form of
\p{totalvpsi-1}. The $\chi^i_A$ part completely decouples and is reduced to the free action. The same decoupling
occurs in the supercharges and the Hamiltonian.
The fermions $\chi^i_A$ contribute in this case only to the $su(2)_R$ generators $J^{ik}$ which uniformly
rotate all doublet indices $i, j, k$.

The Casimir operators of the $su(1,1)$, $su(2)_R$ and $su(2)_L$ subalgebras (at the classical level) defined as
\begin{equation}\label{Cas}
\begin{array}{rcl}
T^2&\equiv& {\textstyle\frac{1}{2}}\, T^{ab}T_{ab} = HK -D^2\,, \\ [6pt]
J^2&\equiv& {\textstyle\frac{1}{2}}\, J^{ik}J_{ik}\,, \\ [6pt]
I^2&\equiv& {\textstyle\frac{1}{2}}\, I^{i^\prime k^\prime}I_{i^\prime k^\prime} = I\bar I
-(I_3)^2
\end{array}
\end{equation}
have the following explicit form
\begin{eqnarray}\label{Cas-11}
T^2&=&
{\textstyle\frac{1}{8}}\,(1+2\alpha) \,
\psi_{i}\psi^{i}\,\bar\psi_{k} \bar\psi^{k}
+{\textstyle\frac{1}{2}}\,\alpha  \,\psi_{i}\bar\psi_{k}\chi^{iA}\chi^{k}_A
-{\textstyle\frac{1}{8}}\,\alpha^2\,\chi_{i}^{A}\chi^{}_{kA}\chi^{iB}\chi^{k}_B\,,  \\ [6pt]
\label{Cas-2}
J^2&=&
{\textstyle\frac{3}{8}}\,
\psi_{i}\psi^{i}\,\bar\psi_{k} \bar\psi^{k}
+{\textstyle\frac{1}{2}}\,\psi_{i}\bar\psi_{k}\chi^{iA}\chi^{k}_A
-{\textstyle\frac{1}{8}}\,\chi_{i}^{A}\chi^{}_{kA}\chi^{iB}\chi^{k}_B\,,
\\ [6pt]
\label{Cas-1}
I^2&=&  -{\textstyle\frac{3}{8}}\,
\psi_{i}\psi^{i}\,\bar\psi_{k} \bar\psi^{k}
\,.
\end{eqnarray}

Using these expressions together with
\begin{equation}\label{QQ-clCas}
{\textstyle\frac{i}{4}}\, Q^{ai^\prime i}Q_{ai^\prime i} =
-{\textstyle\frac{i}{2}}\,(Q^{i}\bar S_{i}- S^{i}\bar Q_{i}) ={\textstyle\frac{1}{2}}\,(1+2\alpha) \,
\psi_{i}\psi^{i}\,\bar\psi_{k} \bar\psi^{k}
+\alpha  \,\psi_{i}\bar\psi_{k}\chi^{iA}\chi^{k}_A\,,
\end{equation}
we find that the second-order (classical) Casimir operator of $D(2,1;\alpha)\,$,
\begin{equation}\label{cl-Cas}
C_2=T^2+\alpha J^2- (1+\alpha)I^2 -{\textstyle\frac{i}{4}}\, Q^{ai^\prime i}Q_{ai^\prime i}\,,
\end{equation}
is expressed as
\begin{equation}\label{cl-Cas-12}
C_2=-{\textstyle\frac{1}{8}}\,\alpha(1+\alpha)\,\chi_{i}^{A}\chi^{}_{kA}\chi^{iB}\chi^{k}_B\,.
\end{equation}
It is worth to point out that the additional fermionic variables $\chi_{i}^{A} $ coming from the multiplet ${\bf (0, 4, 4)}$ make significant contributions
to the $D(2,1;\alpha)$, $su(1,1)$ and $su(2)_R$  Casimirs (\ref{cl-Cas-12}), (\ref{Cas-11}) and (\ref{Cas-2}).

By inspecting the expressions~(\ref{Cas-11})--(\ref{QQ-clCas}), we observe
that, for this particular realization of the $D(2,1;\alpha)$
superalgebra, the following quantity identically vanishes  :
\begin{equation}\label{M-cl}
{M} := {T}^2 - \alpha^2 {J}^2-  {\textstyle\frac{1}{3}}\,(1- \alpha^2)\,{I}^2
-{\textstyle\frac{i}{8}}\, (1-\alpha)\, {Q}^{ai^\prime i} {Q}_{ai^\prime i} =0\,.
\end{equation}
Using this identity together with the expression~(\ref{cl-Cas}), we derive the constraint
\begin{equation}\label{cl-constr}
(1+\alpha)\Big[T^2  - \alpha J^2 +{\textstyle\frac{1}{3}}\,(1-\alpha) I^2\Big]
+(1-\alpha) C_2=0 \,,
\end{equation}
which relates the Casimir of $D(2,1;\alpha)$ to the Casimirs
of the three mutually commuting bosonic subalgebras $su(1,1)$, $su(2)_L$ and $su(2)_R$
in our model. Plugging the expression~(\ref{cl-Cas-12}) for the $D(2,1;\alpha)$ Casimir
into this constraint, we find that
\begin{equation}\label{cl-constr-1}
(1+\alpha)\Big[T^2  - \alpha J^2 +{\textstyle\frac{1}{3}}\,(1-\alpha) I^2
-{\textstyle\frac{1}{8}}\,\alpha(1-\alpha)\,\chi_{i}^{A}\chi^{}_{kA}\chi^{iB}\chi^{k}_B\Big]
=0 \,.
\end{equation}
Using the expressions~(\ref{Cas-11})--(\ref{Cas-1}), we can check that the term
in the square brackets is vanishing, that is the relation
\begin{equation}\label{H-constr-1}
T^2  = \alpha J^2 -{\textstyle\frac{1}{3}}\,(1-\alpha) I^2
+{\textstyle\frac{1}{8}}\,\alpha(1-\alpha)\,\chi_{i}^{A}\chi^{}_{kA}\chi^{iB}\chi^{k}_B
\end{equation}
is valid for any $\alpha\,$. At $\alpha = -1$ and $\alpha = 0$, the Casimir  $C_2$ is vanishing,
while at $\alpha =1$
we have the relation $T^2 = J^2\,$.

Note that the Hamiltonian~(\ref{H-cl}) can be cast in the standard form of the Hamiltonian of (super)conformal
mechanics \cite{FIL2}, \cite{FIL3}, \cite{hkln}
\begin{equation}\label{H-cl-st}
H ={\textstyle\frac{1}{2}}\,p^2  +
\frac{2T^2}{x^2} \,.
\end{equation}
Using the expression~(\ref{H-constr-1}), we can
represent it in the convenient equivalent form
\begin{equation}\label{H-cl-a}
H ={\textstyle\frac{1}{2}}\,p^2  +\alpha(1-\alpha)\,\frac{\chi_{i}^{A}\chi^{}_{kA}\chi^{iB}\chi^{k}_B}{8x^2} + \alpha \,
\frac{J^2}{x^2} - (1-\alpha) \, \frac{I^2}{3x^2}\,.
\end{equation}
The last two terms involve the Casimirs of the groups SU(2)$_R$ and SU(2)$_L$.

Finally, note that the sum of the actions \p{conf_v} and \p{Confpsi} at $\alpha = -\frac13$ is invariant with respect to
a hidden ${\cal N}=8, d=1$ supersymmetry and the exceptional ${\cal N}=8$ superconformal symmetry $F(4)$ \cite{F4scm}. So in this special case
the $D(2,1;\alpha)$ realization given here
 should admit an enlargement to the appropriate realization of $F(4)$. We will not dwell on this point further.

\section{The multiplet pair  $({\bf 3, 4, 1}) \oplus ({\bf 0, 4, 4})$}

\subsection{The multiplet $({\bf 3, 4, 1})$}
This multiplet is described by the analytic superfield $V^{++}(\zeta, u)$ subjected to the off-shell harmonic constraint \cite{IL}
\be \lb{V++}
\begin{array}{c}
D^{++}V^{++} = 0 \\
\Downarrow \\ [5pt]
V^{++} = v^{ik}u^+_iu^+_k + \theta^+\varphi^iu^+_i + \bar\theta^+\bar\varphi^iu^+_i -
i\theta^+\bar\theta^+\left(F + 2\dot{v}^{ik}u^+_iu^-_k\right),
\end{array}
\ee
with all the component fields being functions of $t_A$.

The Grassmann analyticity conditions together with the harmonic constraint \p{V++}, being rewritten in the central basis, imply
\bea
V^{++} = V^{(ik)}(t, \theta, \bar\theta)u^+_iu^+_k\,, \quad D^{(i}V^{kl)} = \bar{D}^{(i}V^{kl)} = 0\,,
\eea
that is solved by
\begin{eqnarray}
V^{(ik)}(t, \theta, \bar\theta)&=& v^{ik} + \theta^{(i}\varphi^{k)} +
\bar\theta^{(i}\bar\varphi^{k)} +
i\theta^{(i}\bar\theta_l \dot v^{k)l}+i\theta_l\bar\theta^{(i} \dot v^{k)l}-i\theta^{(i}\bar\theta^{k)}F \nn
&&
-\,{\textstyle\frac{i}{2}}(\theta)^2\bar\theta^{(i}\dot{\varphi}^{k)} -{\textstyle\frac{i}{2}}(\bar\theta)^2\theta^{(i}\dot{\bar\varphi}{}^{k)} +
{\textstyle\frac{1}{4}}(\theta)^2(\bar\theta)^2 \ddot{v}^{ik}\,,\label{sing-V}
\end{eqnarray}
where $(\overline{v^{ik}})=v_{ik}$, $(\overline{\varphi^i})=\bar\varphi_i$.
The $D(2,1;\alpha)$ transformations of $V^{ik}$, as well as those of the component fields defined in \p{V++}, \p{sing-V}, were given in \cite{IL}.

The superfield $V^{++}$ have the following $D(2,1;\alpha)$
transformation law
\bea
\delta V^{++} = 2\Lambda_{sc} V^{++}\,. \lb{TransfV++}
\eea
The full set of odd $D(2,1;\alpha)$ transformations of the component fields reads
\begin{equation}\label{tr-comp-341}
\begin{array}{c}
\delta v^{ik}=-\omega^{(i}\varphi^{k)}- \bar\omega^{(i}\bar\varphi^{k)}\,, \\ [6pt]
\delta \varphi^i=-2i \bar\omega_k \dot v^{ki} +i \bar\omega^i F -4i\alpha\,\bar\eta_k v^{ki}  \,,
\qquad
\delta \bar\varphi_i= -2i \omega^k \dot v_{ki} -i \omega_i F  -4i\alpha\,\eta^k v_{ki}
\,,\\ [6pt]
\delta F= \omega^k\dot\varphi_{k} +\bar\omega^k \dot{\bar\varphi}_{k}
-(1-2\alpha)\left( \eta^k\varphi_{k} +\bar\eta^k {\bar\varphi}_{k}\right)\,,
\end{array}
\end{equation}
where $\omega_i = \varepsilon_i - t\, \eta_i\, $ and $\bar\omega^i =
\bar\varepsilon^i - t\, \bar\eta^i\, $.

The general sigma-model action of $V^{ik}$ is written as
\be
S_{gen}^{(V)} = \int dtd^4\theta \,L(V) \label{kin1}\ ,
\ee
where $L(V)$ is an arbitrary function of $V^{ik}$. In order to construct the $D(2,1;\alpha)$ invariant subclass of
these actions, we use the explicit expression \p{sing-V} to define
\begin{equation}  \label{sing-V2}
X^{\prime} := \frac{1}{\sqrt{V^2}}= r^\prime + \theta_i\varphi^\prime{}^i +
\bar\varphi^\prime_i\bar\theta^i +
i\theta^{(i}\bar\theta^{k)} A^\prime_{ik}-{\textstyle\frac{i}{2}}(\theta)^2\dot{\varphi}^\prime_i\bar\theta^i
-{\textstyle\frac{i}{2}}(\bar\theta)^2\theta_i\dot{\bar\varphi}{}^\prime{}^i +
{\textstyle\frac{1}{4}}(\theta)^2(\bar\theta)^2 \ddot{r}^\prime\,,
\end{equation}
where
\begin{equation}  \label{1comp-V2}
r^\prime=({v^2})^{-\frac{1}{2}}\,,\qquad
\varphi^\prime{}^i=-(v^2)^{-\frac{3}{2}}\,v^{ik}\varphi_k\,,\qquad
\bar\varphi^\prime_i=(v^2)^{-\frac{3}{2}}\,v_{ik}\bar\varphi^k\,,
\end{equation}
\begin{equation}  \label{1comp-V26}
A^\prime_{ik}=(v^2)^{-\frac{3}{2}}\left[Fv_{ik}-2v_{(i}^l\dot v_{k)l}\right]
\,+\,3i(v^2)^{-\frac{5}{2}}v_{(i}^{j}v_{k)}^{l}\varphi_j\bar\varphi_l
\,+\,{\textstyle\frac{i}{2}}\,(v^2)^{-\frac{3}{2}}\varphi_{(i}\bar\varphi_{k)} \,  .
\end{equation}
This superfield can be checked to transform according to the superconformal transformation law \p{vVconf} of the $({\bf 1, 4, 3})$ multiplet
and obey the constraints \p{Uconstr1} (or, equivalently,  \p{Uconstr2}). It means that the superfield $1/\sqrt{V^2}$ forms
some composite $({\bf 1, 4, 3})$ multiplet.
Therefore, the {\it superconformally invariant} sigma-model type actions of the multiplet $({\bf 3, 4, 1})$ are given by the following
expressions \cite{ikl}
\begin{equation} \label{act-341}
S^{(V)}_{sc} = -{\textstyle\frac{1}{8(1+\alpha)}}\, \int dt d^4\theta\left[
(V^2)^{\frac{1}{2\alpha}}-(V^2)^{-\frac{1}{2}}\right],
\end{equation}
where $V^2 = V^{ik}V_{ik}$. In the limit $\alpha\to -1$ the action \p{act-341} is reduced to
\begin{equation} \label{act-341-1}
S^{(V)(\alpha=-1)}_{sc} = {\textstyle\frac{1}{16}}\,\, \int dt d^4\theta\, (V^2)^{-\frac{1}{2}}
\log V^2 \,.
\end{equation}

The action \p{act-341} has the following component form
\begin{eqnarray}
S^{(V)}_{sc} &=& \textstyle{\frac{1}{8\alpha^2}}\,\displaystyle{\int} dt\,
(v^2)^{\frac{1}{2\alpha}-1}\,\Big[ \dot v^{ik}\dot v_{ik} +\textstyle{\frac{i}{2}}\, \left(\bar\varphi_k \dot\varphi^k
-\dot{\bar\varphi}_k \varphi^k \right) +\textstyle{\frac{1}{2}}\,F^2\, \Big] \nn
&& -\,\textstyle{\frac{i}{8\alpha^2}}\,(\textstyle{\frac{1}{\alpha}}-2)\,\displaystyle{\int} dt\,
(v^2)^{\frac{1}{2\alpha}-2}\,  \Big[ v^{il}\dot v^{k}_l \varphi_{(i}\bar\varphi_{k)}
+\textstyle{\frac{1}{2}}\, F v^{ik}\varphi_{i}\bar\varphi_{k} \Big]
\nonumber \\
&&   - \,
\textstyle{\frac{1}{96\alpha^2}}\,
(\textstyle{\frac{1}{\alpha}}-1)(\textstyle{\frac{1}{\alpha}}-2)\, \displaystyle{\int} dt\,
(v^2)^{\frac{1}{2\alpha}-2}\, \varphi^{i}\bar\varphi^{k} \varphi_{(i}\bar\varphi_{k)} \,.\label{4N-V-c}
\end{eqnarray}
These sigma-model type actions exist at any $\alpha \neq 0$. In what follows we will consider only the $\alpha \neq 0$ options.

Using the analyticity of $V^{++}$, one can construct an off-shell superpotential term for it as an integral over the analytic
subspace
\be \label{SpotV}
S^{(V)}_{sp} = {\textstyle i\frac{\gamma}{\sqrt{2}}} \int du d\zeta^{--} \,L^{++}(V^{++}, u) \,.
\ee
The component form of this action reads
\be \label{half}
{S}_{sp}^{(V)} = {\textstyle\frac{\gamma}{\sqrt{2}}} \int dt \left[F\,{\cal U}(v) +
\dot{v}^{ik}\,{\cal A}_{ik}(v) -i \varphi^{(i}\bar\varphi^{k)}\,\partial_{ik}{\cal U}(v)  \right].
\ee
Here, the background scalar `half-potential' ${\cal U}$ and the magnetic
one-form potential ${\cal A}_{ik}$ are given
by the following harmonic integrals,
\be
{\cal U}(v) = \int du\, \frac{\partial L^{++}}{\partial v^{++}}\,, \qquad
{\cal A}_{ik}(v) = 2 \int du\, u^+_{(i}u^-_{k)}\,
\frac{\partial L^{++}}{\partial v^{++}}\,,\qquad
v^{++} = v^{ik}u^+_iu^+_k\,. \label{VA}
\ee
The genuine scalar potential
$W$ appears as the result of eliminating the auxiliary field $F(t)$ in the sum of the sigma-model
action \p{4N-V-c} and \p{half} as
\be
W(v) = -\sfrac14 \,\gamma^2 \frac{({\cal U}(v))^2}{H(v)}\;.
\label{poten}
\ee
The representation \p{VA} allows one to find the most
general constraints which
${\cal U}$ and ${\cal A}_{ik}$ should obey in order to admit an ${\cal N}{=}4$ supersymmetric
extension:
\bea
&&\partial_{ik}{\cal A}_{lt} - \partial_{lt}{\cal A}_{ik} =
\epsilon_{il}\,\partial_{kt}{\cal U} + \epsilon_{kt}\,\partial_{il}{\cal U}\,,
\qquad\textrm{and}\qquad \Delta {\cal U} =0\,.
\label{monop}
\eea

The $D(2,1;\alpha)$ invariant potential is defined by \cite{IL}
\bea
L^{++}_{sc} = \frac{2\,\hat{V}^{++}}{\sqrt{1 + c^{--}\hat{V}^{++}}
\left(1 + \sqrt{1 + c^{--}\hat{V}^{++}}\right)}\,,  \label{confSpot}
\eea
where
\bea
&&\hat{V}^{++} = V^{++} - c^{++}\,, \quad c^{++} = c^{ik}u^+_iu^+_k\,, \;c^2 = 2\,, \nn
&& \delta_{sc} \hat{V}^{++} = 2\Lambda_{sc}(\hat{V}^{++} + c^{++}) - 2\Lambda_{sc}^{++}c^{+-}\,, \quad
c^{+-} = c^{ik}u^+_{(i}u^+_{k)}\,.\lb{hatTran}
\eea
The analytic Lagrangian \p{confSpot} is invariant under the transformation \p{hatTran} up to a total harmonic derivative.
This can be checked using the variation formula
\bea
\delta_{\hat{V}} L^{++}_{sc} = \frac{\delta \hat{V}^{++}}{(1 + c^{--}\hat{V}^{++})^{3/2}}\,.
\eea
After some algebra, one finds that
\bea
\delta_{sc}L^{++}_{sc} = D^{++} g\,, \quad g = 2\Lambda_{sc} c^{+-} \frac{2 + c^{--}\hat{V}^{++}}{(1 + c^{--}\hat{V}^{++})^{3/2} } -
2\Lambda^{++}_{sc}c^{--}\frac{1}{(1 + c^{--}\hat{V}^{++})^{1/2}}\,.
\eea

The constant triplet $c^{ik}$
breaks spontaneously one of the two mutually commuting $SU(2)$ belonging to $D(2,1;\alpha)$, namely, that one which rotates
the indices $i, k$ of $v^{ik}$ and
$\varphi^i, \bar\varphi^k\,$\footnote{Another $SU(2)$ at any $\alpha$ acts only on fermions, unifying
$\varphi^i$ and $\bar\varphi^k$ into a doublet.}.
This triplet actually parametrizes the Dirac string, as follows from the explicit expressions for
${\cal U}$ and ${\cal A}_{ik}$ in the case under consideration:
\bea
&& {\cal U}^{\,conf} = \int du \frac{1}{\left(\sqrt{1 + c^{--}\hat{v}^{++}}\right)^3}\,,
\label{confpot} \\
&& {\cal A}_{ik}^{\,conf} = 2 \int du\;
\frac{u^+_{(i}u^-_{k)}}{\left(\sqrt{1 + c^{--}\hat{v}^{++}}\right)^3}\,.
\label{Aconf}
\eea
The harmonic integrals in \p{confpot} and \p{Aconf} can be computed to give
\bea
&&{\cal U}^{\,conf} = \frac{\sqrt{2}}{\sqrt{v^{ik}v_{ik}}} =
\frac{|{\bf c}|}{|{\bf v}|}\,,\label{qul} \\
&& {\cal A}_{ik}^{\,conf} = -\sqrt{2}\,
\frac{c_i^pv_{pk} + c_k^pv_{pi}}{\left[({\bf v}\cdot {\bf c}) + |{\bf c}||{\bf v}|\right]|{\bf v}|}\,.
\label{confVect}
\eea
The gauge potential \p{confVect} is transversal
\be \label{tr-pot}
v^{ik}{\cal A}_{ik} =0
\ee
and is recognized as that of Dirac magnetic monopole, with $c^{ik}$ parametrizing
the singular Dirac string. The corresponding field strength computed by eq. \p{monop} does not depend on $c^{ik}$ and
is just the Dirac monopole one
\bea
\partial_{ik}{\cal A}_{lt}^{\,conf} - \partial_{lt}{\cal A}_{ik}^{\,conf} =
-\sqrt{2}\left(\epsilon_{il} v_{kt} + \epsilon_{kt} v_{il}\right)|{\bf v}|^{-3}\,.\lb{monop1}
\eea
Though the magnetic coupling ${\cal A}^{conf} := {\cal A}_{lt}^{conf}\dot{v}^{lt}$ in \p{half} explicitly includes
${\cal A}_{lt}^{conf}$, its $c$-dependence is reduced to  the total $t$-derivative,
as follows from the relation
\bea
c_{(l}^{\,t}\frac{\partial}{\partial c^{m)t}}{\cal A}^{conf} = - \frac{\partial}{\partial t}
\left(\frac{|{\bf c}| v_{lm} + |{\bf v}|c_{lm}}{ ({\bf v}\cdot {\bf c}) + |{\bf c}| |{\bf v}|} \right),
\eea
and so is vanishing under the $t$-integral (up to possible boundary terms).

\subsection{Superconformal coupling of the multiplets $({\bf 3, 4, 1})$ and $({\bf 0, 4, 4})$}

Here we suggest a new way of constructing manifestly ${\cal N}=4, d=1$ conformal coupling of these two multiplets by means of
generating  the superconformal kinetic term of  the nilpotent superfield $\Psi^{+ A}$ through the shift of the analytic superfield $\hat{V}^{++}$
in the superpotential WZ term \p{confSpot}.

Let us define
\bea
W^{++} = \hat{V}^{++} + i\nu\, \Psi^{+A}\Psi^+_A\,.
\eea
This superfield has the same transformation properties as $\hat{V}^{++}$,
\bea
\delta_{sc} W^{++} = 2\Lambda_{sc}(W^{++} + c^{++}) - 2\Lambda^{++}_{sc} c^{+-}\,,
\eea
and, therefore, the substitution of $W^{++}$ for $\hat{V}^{++}$ in \p{confSpot} can not affect the superconformal properties
of this superpotential term. Using the nilpotency property of $\Psi^{+ A}$, the new superconformal WZ term can be written as
\bea
L^{++}_{sc}(W) = L^{++}_{sc}(\hat{V}) + i{\nu}\, \frac{1}{(1 + c^{--}\hat{V}^{++})^{3/2}}\,\Psi^{+A}\Psi^+_A\,.\lb{Shift1}
\eea

Though \p{Shift1} is guaranteed  to be superconformal by construction, it is instructive to explicitly show the invariance
of the second term in \p{Shift1} (up to a total derivative),
\bea
S^{(V,\Psi)}_{sc} = \nu\int du d\zeta^{--} \,L^{++}_{sc}(V,\Psi)\,, \qquad
L^{++}_{sc}(V,\Psi) := \frac{1}{(1 + c^{--}\hat{V}^{++})^{3/2}}\,\Psi^{+A}\Psi^+_A\,. \lb{psivSp}
\eea
Under the superconformal transformations \p{hatTran} and \p{traNPsi1}, the variation of the integrand in \p{psivSp} is reduced, up to a
total harmonic derivative and with making use of the constraint \p{PsiConstr} and \p{V++}, as well as the relation $c^{++}c^{--} - (c^{+-})^2 = 1$, to
\bea
\delta_{sc}L^{++}_{sc}(V,\Psi) = -\frac{\Lambda_{sc}}{(1 + c^{--}\hat{V}^{++})^{3/2}}\Bigg[1 +
3\frac{1 - (c^{+-})^2}{1 + c^{--}\hat{V}^{++} } + 15\frac{(c^{+-})^2}{(1 + c^{--}\hat{V}^{++})^2}\Bigg]\Psi^{+A}\Psi^+_A\,.\lb{VarVpsi}
\eea
After some algebra, denoting $Y := c^{--}\hat{V}^{++}\,$, the right-hand side of this variation can be represented as
\bea
&& \delta_{sc}L^{++}_{sc}(V,\Psi) = (D^{++} f^{--})\,\Psi^{+A}\Psi^+_A\,, \; f^{--} = \Lambda_{sc}c^{--}c^{+-} f_1(Y)
+ \Lambda^{++}_{sc}(c^{--})^2 f_2(Y)\,,  \nn
&& f_1(Y) = - \frac{4 + Y}{(1 + Y)^{5/2}}\,, \quad f_2(Y) = \frac{1}{\sqrt{1 + Y}(1 + Y)}\,.
\eea
Taking into account the constraint \p{PsiConstr}, the variation $\delta_{sc}S^{(V,\Psi)}_{sc}$ vanishes, as expected\footnote{The $(V, \Psi)$
action \p{psivSp} was earlier derived in \cite{BIKL} from a different reasoning.}.

Now we present the off-shell component form of the action in \p{psivSp}
\bea
S^{(V,\Psi)}_{sc} &=& \nu \int dt \Bigg\{2 {\cal U}^{\,conf} \left(-i\phi^{iA}\dot\phi_{iA} + F^A\bar{F}_A \right)
-2i \phi^{iA}\phi_A^{k}\,\dot{v}^{l}_i \partial_{\,kl}\,{\cal U}^{\,conf}   \nn
&& + \,\partial_{kl}{\cal U}^{\,conf}\left[2\left(\varphi^{k}\phi^{lA}\bar{F}_A -\bar\varphi^{k}\phi^{lA}{F}_A\right)
- i \phi^{kA}\phi^{l}_A F \right] \nn
&& - \,\partial_{kl}\partial_{ij}{\cal U}^{\,conf}\,\varphi^{(k}\bar\varphi^{l)}\,\phi^{A(i}\phi^{j)}_A \Bigg\}. \lb{vpsiComp}
\eea
Using the expression \p{qul}, we obtain the explicit form of the component action
\bea
S^{(V,\Psi)}_{sc} &=& b \int dt \,\Bigg\{(v^2)^{-\frac{1}{2}} \left(-i\phi^{iA}\dot\phi_{iA} + F^A\bar{F}_A \right)
-i (v^2)^{-\frac{3}{2}}\,v_i^l\dot{v}_{kl} \,\phi^{iA}\phi_A^{k}   \nn
&& \qquad \qquad - \,(v^2)^{-\frac{3}{2}}\,v_{ik}\,\Big[\left(\varphi^{i}\phi^{kA}\bar{F}_A -\bar\varphi^{i}\phi^{kA}{F}_A\right)
- {\textstyle\frac{i}{2}}\, \phi^{iA}\phi^{k}_A F \Big] \nn
&& \qquad \qquad + \,{\textstyle\frac{1}{2}}\,(v^2)^{-\frac{5}{2}}\,\Big[
(v^2)\,\varphi_{i}\bar\varphi_{k}\,\phi^{iA}\phi^{k}_A-
3\,v_{kl}v_{ij}\,\varphi^{k}\bar\varphi^{l}\,\phi^{iA}\phi^{j}_A
\Big]\,\Bigg\}, \lb{vpsiComp1}
\eea
where
$$
b :=2\sqrt{2}\,\nu \,.
$$

Note that the action \p{vpsiComp1} coincides with the component form of the superfield action
\be \label{Confpsi-341}
S^{(X^\prime, \Psi)}_{sc} =\sfrac12\,b\int du d\zeta^{(-2)}\,{\cal V}^\prime\, \Psi^{+ A}\Psi^{+}_{ A}\,,
\ee
where
\be
{\cal V}^\prime_{WZ}(\zeta, u) = r^\prime(t_A) -2 \theta^+\varphi^\prime{}^i(t_A) u^-_i
-2 \bar\theta^+ \bar\varphi^\prime{}^i(t_A) u^-_i +
3i \theta^+\bar\theta^+ A^\prime{}^{(ik)}(t_A)u^-_iu^-_k \label{WZV1a}
\ee
is the analytic prepotential for the composite $({\bf 1, 4, 3})$ superfield $X^\prime\,$ (cf. \p{WZV1}).

For what follows, it will be useful to explicitly present some intermediate on-shell form of the action \p{vpsiComp1}
by eliminating the auxiliary fields in it. We make use of eq. \p{monop1} and the formulas
\bea
\partial_{kl}{\cal U}^{\,conf} = -\sqrt{2}\,v_{kl}|{\bf v}|^{-3}\,, \quad \partial_{ij}\partial_{kl}{\cal U}^{\,conf}
=\frac{1}{\sqrt{2}}\,|{\bf v}|^{-5}\left[6 v_{ij}v_{kl} - |{\bf v}|^2(\epsilon_{ik}\epsilon_{jl} + \epsilon_{il}\epsilon_{jk}) \right].
\eea
After substituting this into \p{vpsiComp1}, we obtain the following expressions for the auxiliary fields and for the intermediate $(V, \Psi)$
action
\bea
F^A &=& |{\bf v}|^{-2} v_{kl}\,\varphi^{k}\phi^{lA}\,, \qquad \bar F^A = |{\bf v}|^{-2} v_{kl}\,\bar\varphi^{k}\phi^{lA}\,, \lb{aux1}\\
S^{(V,\Psi)}_{sc} &=& b \int dt\, \Big\{-i |{\bf v}|^{-1}\,\phi^{iA}\dot{\phi}_{iA} + {\textstyle\frac{i}{2}}\,|{\bf v}|^{-3}\,
\left[v_{ik} F - 2 v_{i}^{l}\,\dot{v}_{kl}\right]\phi^{iA}\phi^{k}_A \nn
&&\qquad \qquad -\,{\textstyle\frac12}\,|{\bf v}|^{-5}\, v_{ij}v_{kl} \,\varphi^{(i}\bar\varphi^{j)}\,\phi^{A(k}\phi^{l)}_A \,\Big\}. \lb{medVpsi}
\eea

After eliminating the auxiliary field $F$,
\be \label{F-341}
F =-8\alpha^2\gamma (v^2)^{-\frac{1}{2\alpha}+\frac12}+\sfrac{i}{2}\,(\sfrac{1}{\alpha}-2)(v^2)^{-1}\,v_{ik}\varphi^{i}\bar\varphi^{k}
-4ib\alpha^2 (v^2)^{-\frac{1}{2\alpha}-\frac12}\,v_{ik}\phi^{iA}\phi^{k}_A\,,
\ee
in the sum of \p{medVpsi} with the superconformal sigma-model action \p{4N-V-c} and the superpotential WZ action \p{half}
of the $({\bf 3, 4, 1})$ multiplet, we obtain the ultimate on-shell component action of the coupled  $({\bf 3, 4, 1})\oplus ({\bf 0, 4, 4})$
system as
\begin{eqnarray}
S_{(V +\Psi)} &=& \textstyle{\frac{1}{8\alpha^2}}\displaystyle{\int} dt\,
(v^2)^{\frac{1}{2\alpha}-1}\Big[ \dot v^{ik}\dot v_{ik} +\textstyle{\frac{i}{2}}\, \left(\bar\varphi_k \dot\varphi^k
-\dot{\bar\varphi}_k \varphi^k \right)  \Big]
-ib \displaystyle{\int} dt\,  (v^2)^{-\frac{1}{2}}\phi^{iA}\dot{\phi}_{iA} \nonumber\\
&& -\,\textstyle{\frac{i(1-2\alpha)}{8\alpha^3}}\displaystyle{\int} dt\,
(v^2)^{\frac{1}{2\alpha}-2} v_i^{l}\dot v_{kl} \varphi^{(i}\bar\varphi^{k)}
-ib \displaystyle{\int} dt\, (v^2)^{-\frac{3}{2}}
v_{i}^{l}\,\dot{v}_{kl}\phi^{iA}\phi^{k}_A
\nonumber \\
&& +\, {\textstyle\frac{\gamma}{\sqrt{2}}} \int dt  \, \dot{v}^{ik}\,{\cal A}_{ik}(v)\,
-\, 4\alpha^2\gamma^2 \int dt  \, (v^2)^{-\frac{1}{2\alpha}} \nonumber \\
&&   + \,
\textstyle{\frac{i\gamma}{2\alpha}} \displaystyle{\int} dt\,
(v^2)^{-\frac{3}{2}}\, v_{ik}\varphi^{i}\bar\varphi^{k}
- \,
4i\gamma b \alpha^2\displaystyle{\int} dt\,
(v^2)^{-\frac{1}{2\alpha}-1}\,v_{ik} \,\phi^{iA}\phi^{k}_A \nonumber \\
&&   - \,
\textstyle{\frac{1-2\alpha}{192\alpha^4}} \displaystyle{\int} dt\,
(v^2)^{\frac{1}{2\alpha}-2}\, \varphi^{i}\bar\varphi^{k} \varphi_{(i}\bar\varphi_{k)}
- \,
\textstyle{\frac{b}{4\alpha}} \displaystyle{\int} dt\,
(v^2)^{-\frac{5}{2}}\,v_{ij}v_{kl} \,\varphi^{i}\bar\varphi^{j}\,\phi^{Ak}\phi^{l}_A \nonumber \\
&& + \,
\textstyle{\frac{\alpha^2 b^2}{3}} \displaystyle{\int} dt\,
(v^2)^{-\frac{1}{2\alpha}-1}\, \phi^{iA}\phi^{k}_A\,\phi^{B}_i\phi_{kB} \,. \label{4N-V-total}
\end{eqnarray}
At $b=0$ the contribution from the multiplet $({\bf 0,4,4})$ disappears. In what follows we assume that $b\neq 0$.

Introducing the new variables
\begin{equation}\label{4N-341a}
x = (v^2)^{\frac{1}{4\alpha}} \,,
\end{equation}
\begin{equation}\label{4N-341b}
\ell_{ik}={\textstyle\frac{1}{2\alpha}}\,(v^2)^{-\frac{1}{2}}\,v_{ik} \,, \qquad  \ell^{ik}\ell_{ik}= {\textstyle\frac{1}{4\alpha^2}} \,,
\end{equation}
\begin{equation}\label{4N-341c}
\psi^{i} = {\textstyle\frac{1}{2\alpha}}\,(v^2)^{\frac{1}{4\alpha}-1}\, v^{ik}\varphi_{k}\,, \qquad
\bar\psi_{i} = -{\textstyle\frac{1}{2\alpha}}\,(v^2)^{\frac{1}{4\alpha}-1}\, v_{ik}\bar\varphi^{k}\,, \qquad
\chi^{iA} = \sqrt{2b}\,(v^2)^{-\frac{1}{4}}\,\phi^{iA}\,,
\end{equation}
we recast the action \p{4N-V-total} in the form
\begin{eqnarray}
S_{(V+\Psi)} &=& \displaystyle{\int} dt\,\Big\{ \textstyle{\frac{1}{2}}
\left[ \dot x\dot x \, +\,  x^2\, \dot \ell^{ik}\dot \ell_{ik} \,
+ \, i \left(\bar\psi_k \dot\psi^k -\dot{\bar\psi}_k \psi^k  \right) -i\chi^{iA}\dot{\chi}_{iA}\right]
 \nn
&& +\,2i\,\alpha \,
\ell_i^{\,j}\dot \ell_{kj} \Big[ 2\psi^{(i}\bar\psi^{k)} - \alpha\chi^{iA}\chi^{k}_A\Big]
+\, \sqrt{2}\,\alpha\gamma \, \dot{\ell}^{ik}\,{\cal B}_{ik}(\ell)
\nonumber \\
&&
-\, 4\alpha^2\frac{\gamma}{x^{2}} \Big( \gamma\,- \,
2i\, \ell_{ik}\psi^{i}\bar\psi^{k}\,
+ \,i  \alpha\,
\ell_{ik} \,\chi^{iA}\chi^{k}_A \Big) \nonumber \\
&&   - \frac{1}{x^{2}} \Big[\,
\textstyle{\frac{1-2\alpha}{3}} \, \psi^{i}\bar\psi^{k} \psi_{(i}\bar\psi_{k)}
+ \,4\alpha^3 \,
\ell_{ij}\ell_{kl} \,\psi^{i}\bar\psi^{j}\,\chi^{Ak}\chi^{l}_A
- \,
\textstyle{\frac{\alpha^2 }{12}} \, \chi^{iA}\chi^{k}_A\,\chi^{B}_i\chi_{kB} \Big] \Big\}.\label{4N-V-total-341}
\end{eqnarray}

It is worth pointing out that  the gauge potential ${\cal A}_{ik}$ defined in \p{confVect} is now written as
\be
{\cal A}_{ik} (v)= x^{-2\alpha}{\cal B}_{ik} (\ell)\,,
\label{confVect-r}
\ee
where
\be
{\cal B}_{ik} (\ell) = -\frac{4\alpha\, c_{(i}^p \ell_{k)p}}{\sqrt{2}\,\alpha({\bf \ell}\cdot {\bf c}) + 1}\,.
\label{Vect-ell}
\ee
The transversality condition \p{tr-pot} is rewritten as
\be \label{tr-pot-ell}
\ell^{ik}{\cal B}_{ik} =0\,.
\ee

Like in the previous section, one can calculate the classical $D(2,1;\alpha)$ (super)charges and their (anti)commutation relations. At $\alpha=-1$,
the considered system reproduces the relevant $SU(1,1|2)$ invariant on-shell system of Ref. \cite{Galaj}.

The on-shell supersymmetry transformations of the fields in the action \p{4N-V-total-341} are as follows
\begin{eqnarray}\label{3str-X}
\delta x&=&-\omega_i\psi^i+ \bar\omega^i\bar\psi_i\,,
\\ [4pt]
\label{3str-L}
\delta \ell^{ik}&=&-2\alpha\,x^{-1}\left[\omega^{(i}\psi^{j)} +\bar\omega^{(i}\bar\psi^{j)}\right]\ell_j^k
-2\alpha\,x^{-1}\left[\omega^{(k}\psi^{j)} +\bar\omega^{(k}\bar\psi^{j)}\right]\ell_j^i\,,
\\ [4pt]
\delta \psi^i&=&i \bar\omega^i\dot x + i\bar\eta^i x -4\alpha i\, x\,\ell^{m(i}\dot\ell^{k)}_{m}\,\bar\omega_k
-8\alpha^2\gamma i\,x^{-1}\,\ell^{ik}\,\bar\omega_k\nn
&& -\,(1+2\alpha)\,x^{-1}\,\omega_k \psi^{k}\psi^{i}
-2\alpha\,x^{-1}\,\bar\omega_k \psi^{k} \bar\psi^{i}
-8\alpha^2\,x^{-1}\, \ell^{mn}\bar\omega_m \psi_n \,\ell^{ik}\bar\psi_k  \nonumber \\ [3pt]
&& +\,4\alpha^3\,x^{-1}\, \ell^{ik}\bar\omega_k \,\ell_{mn}\chi^{mA} \chi^{n}_A \,,\label{3str-Psi}
\\ [4pt]
\delta \bar\psi_i&=&-i \omega_i\dot x-i\eta_i x -4\alpha i\, x\,\ell^{m}_{(i}\dot\ell_{k)m}\,\omega^k
-8\alpha^2\gamma i\,x^{-1}\,\ell_{ik}\,\omega^k \nn [3pt]
&&  +\,(1+2\alpha)\,x^{-1}\,\bar\omega^k \bar\psi_{k}\bar\psi_{i}
+2\alpha\,x^{-1}\,\omega^k \bar\psi_{k} \psi_{i}
+8\alpha^2\,x^{-1}\, \ell_{mn}\omega^m \bar\psi^n \,\ell_{ik}\psi^k \nonumber \\ [3pt]
&& +\,4\alpha^3\,x^{-1}\, \ell_{ik} \omega^k \,\ell_{mn}\chi^{mA} \chi^{n}_A\,,\label{3str-Psib}
\\ [4pt]
\label{3str-Z}
\delta \chi^{iA}&=&-2\alpha\,x^{-1}\left[\omega^{(i}\psi^{j)} +\bar\omega^{(i}\bar\psi^{j)}\right] \chi^{A}_j\,.
\end{eqnarray}
{}From \p{3str-L} and \p{3str-Z} we observe that the on-shell fermionic transformations of $\ell^{ik}$ and $\chi^{iA}$ have
the form of field-dependent $SU(2)$ transformations like in the case of \p{2str-Z}. The same property is valid for the bosonic semi-dynamical spin variables
entering the ${\cal N}=4$ spin multiplet as the basic ingredient of the ${\cal N}=4$ SCM models of Refs. \cite{FIL2}, \cite{FIL3}.
This resemblance is rather interesting because in our case the angular variables $\ell^{ik}$ are dynamical variables.

Using the component $D(2,1;\alpha)$ transformations \p{3str-X}, \p{3str-L}, \p{3str-Psi}, \p{3str-Psib}, \p{3str-Z},
one can evaluate the corresponding Noether (super)charges
\begin{equation}\label{3Q-cl}
{Q}^i =p\, \psi^i- i x^{-1}\psi_{k}\,\Big[{\textstyle\frac{2}{3}}\,(1+2\alpha)\,\psi^{(i}\bar\psi^{k)}
- \alpha\,\chi^{iA} \chi^{k}_A
-4i\,\alpha\Big(\ell^{i}_p{\cal P}^{kp} -2\,\gamma\alpha\,\ell^{ik}\Big)\Big],
\end{equation}
\begin{equation}\label{3Qb-cl}
\bar{Q}_i=p\, \bar\psi_i+i x^{-1}\bar\psi^{k}\,\Big[{\textstyle\frac{2}{3}}\,(1+2\alpha)\,\psi_{(i}\bar\psi_{k)}
- \alpha\,\chi_i^{A} \chi_{kA}
- 4i\,\alpha\Big(\ell_{ip}{\cal P}_{k}^{p} -2\,\gamma\alpha\,\ell_{ik}\Big)\Big],
\end{equation}
\begin{equation}\label{3S-cl}
{S}^i =x \psi^i - t\,{Q}^i,\qquad \bar{S}_i=x\bar\psi_i-t\,\bar{Q}_i\,.
\end{equation}

Here,  $ p\equiv \dot x$ is the canonical momentum for $x$, and
$$
{\cal P}_{ik} :=p_{ik}-\sqrt{2}\,\alpha\gamma\, x^{2\alpha}{\cal A}_{ik}=p_{ik}-\sqrt{2}\,\alpha\gamma \,{\cal B}_{ik}\,,
$$
with
\begin{equation}\label{mom-l}
p_{ik}=x^2\dot\ell_{ik} + \sqrt{2}\,\alpha\gamma \,{\cal B}_{ik} -
i\alpha \,\ell_i^{\,j}\Big[ 2\psi_{(k}\bar\psi_{j)} - \alpha\chi_k^{A}\chi_{jA}\Big]-
i\alpha \,\ell_k^{\,j}\Big[ 2\psi_{(i}\bar\psi_{j)} - \alpha\chi_i^{A}\chi_{jA}\Big]
\end{equation}
being the momenta for $\ell_{ik}$. The momenta \p{mom-l} satisfy the constraint
\begin{equation}\label{costr-mom-l}
\ell^{ik}p_{ik}=0\,,
\end{equation}
which forms the pair of the second class constraints together with the constraint \p{4N-341b},
\begin{equation}\label{costr-ell-l}
\ell^{ik}\ell_{ik}-{\textstyle\frac{1}{4\alpha^2}}=0 \,.
\end{equation}
Then one introduces Dirac brackets for the second class constraints \p{costr-mom-l}, \p{costr-ell-l}
\begin{equation}\label{DB-l}
\begin{array}{c}
\{ \ell^{ij}, \ell^{\,km} \}_{{}_{DB}}= 0\,,\qquad
\{ \ell^{ij}, p_{\,km} \}_{{}_{DB}}= \delta_{(k}^{i}\delta_{m)}^{j} -4\alpha^2\ell^{ij}\ell_{\,km}\,, \\ [7pt]
\{ p_{ij}, p_{\,km} \}_{{}_{DB}}= 4\alpha^2\left( p_{ij} \ell_{\,km} - p_{\,km} \ell_{ij}\right).
\end{array}
\end{equation}

Using these Dirac  brackets, one can check that the supercharges \p{3Q-cl}, \p{3Qb-cl}, \p{3S-cl} form $D(2,1;\alpha)$ algebra \p{DB-QQ}
with the Hamiltonian
\begin{eqnarray}
H &=&{\displaystyle{\textstyle\frac{1}{2}}\,p^2  + \frac{1}{x^{2}}\Bigg[\textstyle\frac{1}{2}}\,{\cal P}^{ia}{\cal P}_{ia}
+ 4\alpha^2\gamma^2 +\,2i\alpha \Big(\ell^{i}_{m} {\cal P}^{\,km}-2\gamma\alpha\,\ell^{ik}\Big)\Big( 2\psi_{(i}\bar\psi_{k)} - \alpha\chi_i^{A}\chi_{kA}\Big)\nn
&&\qquad\qquad\quad +\,{\textstyle\frac{1}{4}}\,(1+2\alpha) \,
\psi_{i}\psi^{i}\,\bar\psi_{k} \bar\psi^{k}
+\alpha  \,\psi_{i}\bar\psi_{k}\chi^{iA}\chi^{k}_A
-{\textstyle\frac{1}{4}}\,\alpha^2\,\chi_{i}^{A}\chi^{}_{kA}\chi^{iB}\chi^{k}_B \Bigg]\label{H-cl-341}
\end{eqnarray}
and SU(2) generators
\begin{equation} \label{J-cl-341}
J^{ik} = -i\Big[ \psi^{(i}\bar\psi{}^{k)}-{\textstyle\frac12}\,\chi^{iA}\chi^k_{A}-2i\Big(\ell^{i}_{m} {\cal P}^{\,km}-2\gamma\alpha\,\ell^{ik}\Big)\Big].
\end{equation}
The remaining even generators $K$, $D$ and $I^{i^\prime k^\prime}$ are given by the same expressions as in \p{K-cl}, \p{D-cl} and \p{I-cl}, \p{I-cl1}.

Note the useful relations
\begin{equation} \label{id-l-341}
\ell^{\,i}_{\,m} p^{\,km}=\ell^{\,(i}_{\,m} p^{\,k)m}\,, \qquad
\ell^{\,i}_{\,m} p^{\,km}\, \ell_{ij} p_{\,k}^{\,j}={\textstyle\frac{1}{8\alpha^2}}\,p^{\,km}p_{\,km} \,,
\end{equation}
\begin{equation} \label{id-2-341}
\Big(\ell^{i}_{m} {\cal P}^{\,km}-2\gamma\alpha\,\ell^{ik}\Big)\Big(\ell_{ij} {\cal P}^{\,j}_{\,k}-
2\gamma\alpha\,\ell_{ik}\Big)={\textstyle\frac{1}{8\alpha^2}}\,\Big({\textstyle\frac{1}{2}}\,{\cal P}^{ia}{\cal P}_{ia}
+\, 4\alpha^2\gamma^2\Big),
\end{equation}
\begin{equation} \label{id-3-341}
\ell^{i}_{m} {\cal P}^{\,km}-2\gamma\alpha\,\ell^{ik}=\ell^{i}_{m} {p}^{\,km} -\gamma\,
\frac{2\alpha|{\bf c}|\ell^{ik}+c^{ik}}{2\alpha({\bf \ell}\cdot{\bf c})+|{\bf c}|}\,,
\end{equation}
which follow from the definitions of $\ell^{ik}, {\cal P}_{ik}$ and $p_{ik}$.
Using these relations, we can represent the Casimir operators \p{Cas} of the $su(1,1)$, $su(2)_R$ and $su(2)_L$ subalgebras as
\begin{eqnarray}
T^2&=&
{\textstyle\frac{1}{8}}\,(1+2\alpha) \,
\psi_{i}\psi^{i}\,\bar\psi_{k} \bar\psi^{k}
+\textstyle\frac{1}{2}\,\alpha  \,\psi_{i}\bar\psi_{k}\left[\chi^{iA}\chi^{k}_A +4i\,\Big(\ell^{i}_{m} {\cal P}^{\,km}-2\gamma\alpha\,\ell^{ik}\Big)\right]
\nn [6pt]
&&
-\,{\textstyle\frac{\alpha^2}{8}}\left[\chi_{i}^{A}\chi^{}_{kA}+4i\,\Big(\ell_{im} {\cal P}_{\,k}^{\,m}-2\gamma\alpha\,\ell_{ik}\Big)\right]
\left[\chi^{iB}\chi^{k}_B +4i\,\Big(\ell^{i}_{p} {\cal P}^{\,kp}-2\gamma\alpha\,\ell^{ik}\Big)\right],  \label{Cas-11-341}\\
J^2 &=&
{\textstyle\frac{3}{8}}\,
\psi_{i}\psi^{i}\,\bar\psi_{k} \bar\psi^{k}
+{\textstyle\frac{1}{2}}\,\psi_{i}\bar\psi_{k}\left[\chi^{iA}\chi^{k}_A +4i\,\Big(\ell^{i}_{m} {\cal P}^{\,km}-2\gamma\alpha\,\ell^{ik}\Big)\right] \nn [6pt]
&&
-{\textstyle\frac{1}{8}}\left[\chi_{i}^{A}\chi^{}_{kA}+4i\,\Big(\ell_{im} {\cal P}_{\,k}^{\,m}-2\gamma\alpha\,\ell_{ik}\Big)\right]
\left[\chi^{iB}\chi^{k}_B +4i\,\Big(\ell^{i}_{p} {\cal P}^{\,kp}-2\gamma\alpha\,\ell^{ik}\Big)\right],
\label{Cas-2-341} \\
I^2&=&  -{\textstyle\frac{3}{8}}\,
\psi_{i}\psi^{i}\,\bar\psi_{k} \bar\psi^{k}
\,.\label{Cas-1-341}
\end{eqnarray}
Exploiting these expressions and the relation
\begin{equation}\label{QQ-clCas-341}
{\textstyle\frac{i}{4}}\, Q^{ai^\prime i}Q_{ai^\prime i} ={\textstyle\frac{1}{2}}\,(1+2\alpha) \,
\psi_{i}\psi^{i}\,\bar\psi_{k} \bar\psi^{k}
+\alpha  \,\psi_{i}\bar\psi_{k}\left[\chi^{iA}\chi^{k}_A +4i\,\Big(\ell^{i}_{m} {\cal P}^{\,km}-2\gamma\alpha\,\ell^{ik}\Big)\right],
\end{equation}
we obtain the second-order (classical) Casimir operator of $D(2,1;\alpha)\,$ in the form
\begin{eqnarray}\label{cl-Cas-12-341}
C_2=-{\textstyle\frac{1}{8}}\,\alpha(1+\alpha)\left[\chi_{i}^{A}\chi^{}_{kA}+4i\,\Big(\ell_{im} {\cal P}_{\,k}^{\,m}-2\gamma\alpha\,\ell_{ik}\Big)\right]
\Big[\chi^{iB}\chi^{k}_B +4i\,\Big(\ell^{i}_{p} {\cal P}^{\,kp}-2\gamma\alpha\,\ell^{ik}\Big)\Big].
\end{eqnarray}
Note that, as in the case of the pair $({\bf 1, 4, 3})\oplus ({\bf 0, 4, 4})$, the additional fermionic variables
$\chi_{i}^{A} $ make significant contributions to the Casimirs (\ref{Cas-11-341}), (\ref{Cas-2-341}) and (\ref{cl-Cas-12-341}).

Like in the previous case, looking at the expressions~(\ref{Cas-11-341})--(\ref{QQ-clCas-341}), we observe
that the following quantity $M$ vanishes identically for this particular realization of the $D(2,1;\alpha)$
superalgebra:
\begin{equation}\label{M-cl-341}
{M}\equiv {T}^2 - \alpha^2 {J}^2-  {\textstyle\frac{1}{3}}\,(1- \alpha^2)\,{I}^2
-{\textstyle\frac{i}{8}}\, (1-\alpha)\, {Q}^{ai^\prime i} {Q}_{ai^\prime i} =0\,.
\end{equation}
Taking into account the expression~(\ref{QQ-clCas-341}), this identity implies the constraint
\begin{equation}\label{cl-constr-341}
(1+\alpha)\Big[T^2  - \alpha J^2 +{\textstyle\frac{1}{3}}\,(1-\alpha) I^2\Big]
+(1-\alpha) C_2=0 \,,
\end{equation}
which relates the Casimir of $D(2,1;\alpha)$ to the Casimirs
of the three mutually commuting bosonic subalgebras $su(1,1), su(2)_L$ and $su(2)_R$
in our model. Plugging the expression~(\ref{cl-Cas-12-341}) for the $D(2,1;\alpha)$ Casimir
into this constraint, we find (cf. \p{cl-constr-1}):
\begin{eqnarray}
&&(1+\alpha)\Big\{T^2  - \alpha J^2 +{\textstyle\frac{1}{3}}\,(1-\alpha) I^2
-{\textstyle\frac{1}{8}}\,\alpha(1-\alpha)\left[\chi_{i}^{A}\chi^{}_{kA}+4i\,\Big(\ell_{im} {\cal P}_{\,k}^{\,m}-2\gamma\alpha\,\ell_{ik}\Big)\right]\times \nn
&&
\times \left[\chi^{iB}\chi^{k}_B +4i\,\Big(\ell^{i}_{p} {\cal P}^{\,kp}-2\gamma\alpha\,\ell^{ik}\Big)\right]\Big\}
=0 \,.\label{cl-constr-1-341}
\nonumber
\end{eqnarray}
Using the expressions~(\ref{Cas-11-341})--(\ref{Cas-1-341}), we can check that the expression
within the curled brackets is vanishing on its own, whence
\begin{eqnarray}
T^2  &= & \alpha J^2 -{\textstyle\frac{1}{3}}\,(1-\alpha) I^2
+{\textstyle\frac{1}{8}}\,\alpha(1-\alpha)\left[\chi_{i}^{A}\chi^{}_{kA}+4i\,\Big(\ell_{im} {\cal P}_{\,k}^{\,m}-2\gamma\alpha\,\ell_{ik}\Big)\right]\times \nn
&&\times \left[\chi^{iB}\chi^{k}_B +4i\,\Big(\ell^{i}_{p} {\cal P}^{\,kp}-2\gamma\alpha\,\ell^{ik}\Big)\right].\label{H-constr-1-341}
\nonumber
\end{eqnarray}

Note that the Hamiltonian~(\ref{H-cl-341}) has the standard form of the Hamiltonian of (super)conformal mechanics
\begin{equation}\label{H-cl-st-341}
H ={\textstyle\frac{1}{2}}\,p^2  +
\frac{2T^2}{x^2} \,.
\end{equation}
Using the relation~(\ref{H-constr-1-341}), we can
rewrite the Hamiltonian in the form
\begin{eqnarray}
H &=& {\textstyle\frac{1}{2}}\,p^2   + \alpha \,
\frac{J^2}{x^2} - (1-\alpha) \, \frac{I^2}{3x^2} +
\frac{\alpha(1-\alpha)}{8x^2}\Big[\chi_{i}^{A}\chi^{}_{kA}+4i\,\Big(\ell_{im} {\cal P}_{\,k}^{\,m}-2\gamma\alpha\,\ell_{ik}\Big)\Big]\times \nn
&&
\times \,\Big[\chi^{iB}\chi^{k}_B +4i\,\Big(\ell^{i}_{p} {\cal P}^{\,kp}-2\gamma\alpha\,\ell^{ik}\Big)\Big].\label{H-cl-a-341}
\end{eqnarray}
This representation is analogous to \p{H-cl-a}.

\section{The multiplet pair  $({\bf 4, 4, 0}) \oplus ({\bf 0, 4, 4})$}

\subsection{The multiplet $({\bf 4, 4, 0})$}
This multiplet is described by the analytic superfield $q^{+}_a(\zeta, u)\,, \,\widetilde{q^+_a} = \epsilon^{ab}q^+_b := q^{+a}\,, \,
\widetilde{q^{+a}} = - q^+_a\,,$ subjected to the off-shell harmonic constraint \cite{IL}
\be \lb{q+}
\begin{array}{c}
D^{++}q^{+ a} = 0 \\
\Downarrow \\ [5pt]
q^{+ a} = f^{i a}u^+_i  + \theta^+\varphi^a + \bar\theta^+\bar\varphi^a
- 2i\theta^+\bar\theta^+ \dot{f}^{ia}u^-_i\,.
\end{array}
\ee
Here $a = 1,2$ is the doublet index of some extra ``Pauli-G\"ursey'' $SU(2)$ which commutes with both the Poincar\'e and conformal
${\cal N}{=}4, d{=}1$ supersymmetries. The harmonic constraint \p{q+} together with the analyticity conditions amount in the central basis
to \cite{ikl1}
\bea
q^{+ a} = q^{ia}(t, \theta, \bar\theta)u^+_i\,, \quad D^{(i}q^{k)a} = \bar{D}^{(i}q^{k)a} = 0\,, \lb{qcentr}
\eea
where
\begin{equation}  \label{central-q}
q^{ia}(t,\theta_i,\bar\theta^i)= f^{i a} + \theta^i\varphi^a +
\bar\theta^i \bar\varphi^a -
2i\,\theta^{(i}\bar\theta^{k)} \dot f^{a}_{k}-{\textstyle\frac{i}{2}}(\theta)^2\bar\theta^i\dot{\varphi}^a
-{\textstyle\frac{i}{2}}(\bar\theta)^2\theta^i\dot{\bar\varphi}{}^a +
{\textstyle\frac{1}{4}}(\theta)^2(\bar\theta)^2 \ddot f^{i a}\,.
\end{equation}

Under the superconformal symmetry the analytic superfield $q^{+ a}$ transforms as
\bea
\delta_{sc} q^{+ a} = \Lambda_{sc} q^{+ a}\,.
\eea
The full set of the supersymmetry transformations acts on the component fields as
\begin{equation}\label{tr-comp-440}
\begin{array}{c}
\delta f^{ia}=-\omega^{i}\varphi^{a}- \bar\omega^{i}\bar\varphi^{a}\,, \\ [6pt]
\delta \varphi^a=-2i \bar\omega_k \dot f^{ka} -2i\alpha\,\bar\eta_k f^{ka}  \,,
\qquad
\delta \bar\varphi_a= -2i \omega^k \dot f_{ka} -2i\alpha\,\eta^k f_{ka}
\,,
\end{array}
\end{equation}
where $\omega_i = \varepsilon_i - t\, \eta_i\, $ and $\bar\omega^i =
\bar\varepsilon^i - t\, \bar\eta^i\, $.

The general sigma-model type $q^+$-action reads
\be
S^{(q)}_{gen} = \int dtd^4\theta\, L(q^{ia})\,,
\ee
with $L(q)$ being an arbitrary function of $q^{ia}$.

Using the explicit component expansion \p{central-q} and defining $(q^2)^{-1}\equiv (q^{ia}q_{ia})^{-1}\,$, we can construct the composite
superfield
\begin{equation}  \label{sing-q2}
X^{\prime\prime}\equiv\frac{1}{q^2}= r^{\prime\prime} + \theta_i\varphi^{\prime\prime}{}^i +
\bar\varphi^{\prime\prime}_i\bar\theta^i +
i\theta^{(i}\bar\theta^{k)} A^{\prime\prime}_{ik}-{\textstyle\frac{i}{2}}(\theta)^2\dot{\varphi}^{\prime\prime}_i\bar\theta^i
-{\textstyle\frac{i}{2}}(\bar\theta)^2\theta_i\dot{\bar\varphi}{}^{\prime\prime}{}^i +
{\textstyle\frac{1}{4}}(\theta)^2(\bar\theta)^2 \ddot{r}^{\prime\prime}\,,
\end{equation}
where
\begin{equation}  \label{1comp-r2}
r^{\prime\prime}=(f^2)^{-1}\,,\qquad
\varphi^{\prime\prime}{}^i=-2(f^2)^{-2}\,f^{ia}\varphi_a\,,\qquad
\bar\varphi^{\prime\prime}_i=2(f^2)^{-2}\,f_{ia}\bar\varphi^a\,,
\end{equation}
\begin{equation}  \label{1comp-A2}
A^{\prime\prime}_{ik}=-4(f^2)^{-2}f_{(i}^a\dot f_{k)a}
\,+\,8i(f^2)^{-3}f_{(i}^{a}f_{k)}^{b}\varphi_a\bar\varphi_b
\end{equation}
and $f^2\equiv f^{ia}f_{ia}$. One can check that $X^{\prime\prime}$ transforms under $D(2,1;\alpha)$ as the standard $({\bf 1, 4, 3})$
superfield $X$ and obeys the same constraints.  Hence, the superfield $1/q^2$ represents a composite $({\bf 1, 4, 3})$ multiplet.
Therefore, the {\it superconformally invariant} sigma-model type actions of the multiplet $({\bf 4, 4, 0})$ can be constructed just on
the pattern of the $X$ superconformal action \p{conf_v}:
\begin{equation} \label{act-440}
S^{(q)}_{sc} = -{\textstyle\frac{1}{8(1+\alpha)}}\, \int dt d^4\theta\left[
(q^2)^{\frac{1}{\alpha}}-(q^2)^{-1}\right],
\end{equation}
where $q^2 = q^{ia}q_{ia}$. In the limit $\alpha\to -1$ the action \p{act-440} goes over to
\begin{equation} \label{act-440-1}
S^{(q)(\alpha=-1)}_{sc} = {\textstyle\frac{1}{8}}\, \int dt d^4\theta\, (q^2)^{-1}
\log q^2\,.
\end{equation}

The action \p{act-440} has the following component formulation
\begin{eqnarray}
S^{(q)}_{sc} &=& \textstyle{\frac{1}{2\alpha^2}}\,\displaystyle{\int} dt\,
(f^2)^{\frac{1}{\alpha}-1}\,\Big[ \dot f^{ia}\dot f_{ia} +\textstyle{\frac{i}{2}}\, \left(\bar\varphi_a \dot\varphi^a
-\dot{\bar\varphi}_a \varphi^a \right) \Big] \nn
&& -\,\textstyle{\frac{i}{\alpha^2}}\,(\textstyle{\frac{1}{\alpha}}-1)\,\displaystyle{\int} dt\,
(f^2)^{\frac{1}{\alpha}-2}\,   f^{ia}\dot f^{b}_i \varphi_{(a}\bar\varphi_{b)}
\nonumber \\
&&   - \,
\textstyle{\frac{1}{6\alpha^3}}\,
(\textstyle{\frac{1}{\alpha}}-1)\, \displaystyle{\int} dt\,
(f^2)^{\frac{1}{\alpha}-2}\, \varphi^{a}\bar\varphi^{b} \varphi_{(a}\bar\varphi_{b)} \,.\label{4N-440-c}
\end{eqnarray}

The component bosonic actions is meaningful  only under the assumption that the `vacuum value'
of the radial part of $q^{ia}$ is non-vanishing, i.e. $\langle q^2\rangle\neq 0$.
It is worth noting that the extra Pauli-G\"ursey SU(2) group acting on the index
$a$ of $q^{ia}$ is respected by the superconformal actions \p{act-440}, \p{4N-440-c}.
Thus these actions are manifestly O(4) invariant.

For the $({\bf 4, 4, 0})$ multiplet one can also define the superpotential WZ term
\be
S_{sp}^{(q)} = -\sfrac{i}{2} \int du d\zeta^{(-2)} {\cal L}^{++} (q^+, u)\,, \lb{WZq}
\ee
which in components yields the coupling $\sim \dot{f}^{ia}{\cal A}_{ia}$, as in the case of
the $({\bf 3, 4, 1})$ multiplet. However, as opposed to the latter case, the superconformal subclass of \p{WZq} is vanishing
because the corresponding component Lagrangian is reduced to a total $t$-derivative. This conclusion is based
on the following reasoning. Define
\be
{\cal V}^{++} = q^{+ a}a_{ab}q^{+b}\,, \quad a^{ab}a_{ab} = 2\,, \label{composite}
\ee
where $a_{ab} = a_{ba}$ is a constant triplet which breaks the extra Pauli-G\"ursey SU(2) (realized on the indices $a, b$) down to
its some U(1) subgroup. The composite analytic superfield
${\cal V}^{++}$ has the same transformation rule under $D(2,1;\alpha)$ as $V^{++}\,$, $\delta {\cal V}^{++} = 2\Lambda_{sc}{\cal V}^{++}\,$,
so one can apply the same method of constructing a superconformal WZ term for it. After substituting the component fields of ${\cal V}^{++}$,
\be
{\cal V}_0^{ik} = f^{ia}a_{ab}f^{kb}\,, \; {\cal F} = 2\,(\,\dot{f}^{\;a}_i a_{ab} f^{ib}
+ i\varphi^a a_{ab}\bar \varphi^b\,)\,, \;
\chi^i = 2 \varphi^aa_{ab}f^{ib}\,, \; \bar \chi^i =
2 \bar\varphi^a a_{ab} f^{ib}\,,\label{compcompos}
\ee
into \p{half}, the fermionic contributions from the first and third terms in  \p{half} cancel each other, while the bosonic rest is reduced to
\be
S^{(q)}_{sp\ (sc)}\;\;\Longrightarrow \;\;
\int dt \dot{f}^{ia}{\cal A}_{ia}^{conf}(f)\,, \label{confAf}
\ee
where
\be
{\cal A}^{conf}_{ia}(f) =
\frac{4}{\left[f^2 + (f\cdot a\cdot f\cdot c)\right]}\left(f^k_ac_{ki}
- f^b_ia_{ab} \right)
\label{potq}
\ee
and
$$
f\cdot a\cdot f\cdot c = f^{ia}f^{kb}c_{ik}a_{ab}\,.
$$
Calculating the curl of the vector potential \p{potq}, we find
\be
\partial_{kb}{\cal A}^{conf}_{ia} - \partial_{ia}{\cal A}^{conf}_{kb} = 0\,,\lb{zerocurv}
\ee
i.e. ${\cal A}^{conf}_{ia}$ is a pure gauge and the Lagrangian in \p{confAf} is a total time derivative.

\subsection{Superconformal couplings of the multiplets $({\bf 4, 4, 0})$ and $({\bf 0, 4, 4})$}

Similarly to \p{WZV1}, we can construct the ``prepotential'' for the superfield $1/q^2$:
\be
{\cal V}^{\,\prime\prime}_{WZ}(\zeta, u) = r^{\prime\prime}(t_A) -2 \theta^+\varphi^{\,\prime\prime}{}^i(t_A) u^-_i
-2 \bar\theta^+ \bar\varphi^{\prime\prime}{}^i(t_A) u^-_i +
3i \theta^+\bar\theta^+ A^{\prime\prime}{}^{(ik)}(t_A)u^-_iu^-_k\,, \label{WZV1b}
\ee
where $r^{\prime\prime}$, $\varphi^{\prime\prime}{}^i$, $\bar\varphi^{\prime\prime}{}^i$ and $A^{\prime\prime}{}^{(ik)}$
are defined in \p{1comp-r2}, \p{1comp-A2}.

Then the  superconformal coupling is given by the superfield action
\be \label{Confpsi-440}
S^{(X^{\,\prime\prime}, \Psi)}_{sc} =\sfrac12\,b\int du d\zeta^{(-2)}\,{\cal V}^{\,\prime\prime}\, \Psi^{+ A}\Psi^{+}_{ A}\,.
\ee

It is easy to specialize the component superconformal action \p{coup-143-comp} to this case:
\bea \lb{vpsiComp2}
S^{(q,\Psi)}_{sc} &=& b \int dt \,\Bigg\{(f^2)^{-1} \left(-i\phi^{iA}\dot\phi_{iA} + F^A\bar{F}_A \right)
-2i (f^2)^{-2}\,f_i^a\dot{f}_{ka} \,\phi^{iA}\phi_A^{k}   \nn
&& \qquad \qquad - \,2(f^2)^{-2}\,f_{ka}\,\Big(\varphi^{a}\phi^{kA}\bar{F}_A -\bar\varphi^{a}\phi^{kA}{F}_A\Big) \nn
&& \qquad \qquad - \, 4\,(f^2)^{-3}f_{ia}f_{kb}\,\varphi^{a}\bar\varphi^{b}\,\phi^{iA}\phi^{k}_A \,\Bigg\}.
\eea
Elimination of the auxiliary fields $F^A, \bar{F}_A$  by their equations of motion,
\be
F^A= 2(f^2)^{-1}f_{ka}\varphi^{a}\phi^{kA}\,,\qquad \bar{F}_A =2(f^2)^{-1}f^{ka}\bar\varphi_{a}\phi_{kA}\,,
\ee
nullifies the total four-fermionic term in \p{vpsiComp2}, resulting in the very simple on-shell action
\be \lb{vpsiComp2-a}
S^{(q,\Psi)}_{sc} = -i b \int dt \, (f^2)^{-1} \left[ \phi^{iA}\dot\phi_{iA}
+ 2(f^2)^{-1}\,f_i^a\dot{f}_{ka} \,\phi^{iA}\phi_A^{k}   \right].
\ee

The same superconformal  coupling can be constructed just through the substitution $\hat{V}^{++} \rightarrow {\cal V}^{++}$ in \p{Shift1}, where
the composite analytic superfield ${\cal V}^{++}$ was defined in \p{composite}. While $L^{++}_{sc}({\cal V})$ is reduced to the total derivative and
so does not contribute, the rest of \p{Shift1} is a  non-trivial $D(2,1;\alpha)$ invariant off-shell coupling of the multiplets
$({\bf 0, 4, 4})$ and $({\bf 4, 4, 0})$:
\bea
S^{(q,\Psi)}_{sc} = b\int du d\zeta^{--} \,L^{++}_{sc}(q,\Psi)\,, \qquad
L^{++}_{sc}(q,\Psi) := \frac{1}{(1 + c^{--}\hat{\cal V}^{++})^{3/2}}\,\Psi^{+A}\Psi^+_A\,. \lb{psiqSp}
\eea
In components, it yields the same actions \p{vpsiComp2} and \p{vpsiComp2-a}.

The total on-shell superconformal action is the sum of the component superconformal ${\bf (4,4,0)}$ action \p{4N-440-c} and
the action \p{vpsiComp2-a}:
\begin{eqnarray}
S_{(q+\Psi)} &=& \textstyle{\frac{1}{2\alpha^2}}\,\displaystyle{\int} dt\,
(f^2)^{\frac{1}{\alpha}-1}\,\Big[ \dot f^{ia}\dot f_{ia} +\textstyle{\frac{i}{2}}\, \left(\bar\varphi_a \dot\varphi^a
-\dot{\bar\varphi}_a \varphi^a \right) \Big] -i b \displaystyle{\int} dt \, (f^2)^{-1}  \phi^{iA}\dot\phi_{iA} \nonumber\\
&& -\,\textstyle{\frac{i}{\alpha^2}}\,(\textstyle{\frac{1}{\alpha}}-1)\,\displaystyle{\int} dt\,
(f^2)^{\frac{1}{\alpha}-2}\,   f^{ia}\dot f^{b}_i \varphi_{(a}\bar\varphi_{b)}
-2i b \int dt \, (f^2)^{-2} f_i^a\dot{f}_{ka} \,\phi^{iA}\phi_A^{k}
\nonumber \\
&&   - \,\textstyle{\frac{1}{6\alpha^3}}\,
(\textstyle{\frac{1}{\alpha}}-1)\, \displaystyle{\int} dt\,
(f^2)^{\frac{1}{\alpha}-2}\, \varphi^{a}\bar\varphi^{b} \varphi_{(a}\bar\varphi_{b)} \,.\label{4N-440-t}
\end{eqnarray}
Thus in this case fermionic fields from different multiplets interact only with bosonic fields
and the four-fermionic term is composed only out of the fermionic fields of the ${\bf (4,4,0)}$ multiplet.

Introducing the new variables
\begin{equation}\label{4N-440a}
x = (f^2)^{\frac{1}{2\alpha}} \,,
\end{equation}
\begin{equation}\label{4N-440b}
\textsc{l}_{ia}={\textstyle\frac{1}{\alpha}}\,(f^2)^{-\frac{1}{2}}\,f_{ia} \,, \qquad  \textsc{l}^{ia}\textsc{l}_{ia}
= {\textstyle\frac{1}{\alpha^2}} \,,
\end{equation}
\begin{equation}\label{4N-440c}
\psi^{i} = {\textstyle\frac{1}{\alpha}}\,(f^2)^{\frac{1}{2\alpha}-1}\, f^{ia}\varphi_{a}\,, \qquad
\bar\psi_{i} = -{\textstyle\frac{1}{\alpha}}\,(f^2)^{\frac{1}{2\alpha}-1}\, f_{ia}\bar\varphi^{a}\,, \qquad
\chi^{iA} = \sqrt{2b}\,(f^2)^{-\frac{1}{2}}\,\phi^{iA}\,,
\end{equation}
at $\alpha \neq 0$ we can rewrite the action \p{vpsiComp2-a} in the form
\begin{eqnarray}
S_{(q+\Psi)} &=& \displaystyle{\int} dt\,\Bigg\{ \textstyle{\frac{1}{2}}
\left[ \dot x\dot x \, +\,  x^2\, \dot{\textsc{l}}^{ia}\dot{\textsc{l}}_{ia} \,
+ \, i \left(\bar\psi_k \dot\psi^k -\dot{\bar\psi}_k \psi^k  \right) -i\chi^{iA}\dot{\chi}_{iA}\right]
 \nn
&& +\,i\,\alpha \,
{\textsc{l}}_i^{\,a}\dot{\textsc{l}}_{ka} \Big[ 2\psi^{(i}\bar\psi^{k)} - \alpha\chi^{iA}\chi^{k}_A\Big]
-
\textstyle{\frac{2(1-\alpha)}{3}} \, x^{-2} \,\psi^{i}\bar\psi^{k} \psi_{(i}\bar\psi_{k)}
\Bigg\} \,.\label{4N-V-total-440}
\end{eqnarray}

The on-shell supersymmetry transformations leaving the action \p{4N-V-total-440} invariant are
\begin{eqnarray}\label{4str-X}
\delta x&=&-\omega_i\psi^i+ \bar\omega^i\bar\psi_i\,,
\\ [4pt]
\label{4str-L}
\delta \textsc{l}^{ia}&=&-2\alpha\,x^{-1}\left[\omega^{(i}\psi^{j)} +\bar\omega^{(i}\bar\psi^{j)}\right]\textsc{l}_j{}^a
\,,
\\ [4pt]
\delta \psi^i&=&i \bar\omega^i\dot x + i\bar\eta^i x -2\alpha i\, x\,\textsc{l}^{ia}\dot{\textsc{l}}^{k}{}_{a}\,\bar\omega_k \nn [3pt]
&& -(1+2\alpha)\,x^{-1}\,\omega_k \psi^{k}\psi^{i}
-2\alpha\,x^{-1}\,\bar\omega_k \psi^{k} \bar\psi^{i}
+x^{-1}\, \bar\omega_k \psi^i \bar\psi^k   \,, \label{4str-Psi}
\\ [4pt]
\delta \bar\psi_i&=&-i \omega_i\dot x-i\eta_i x -2\alpha i\, x\,\textsc{l}_{i}{}^{a}\dot{\textsc{l}}_{ka}\,\omega^k \nn [3pt]
&&  +(1+2\alpha)\,x^{-1}\,\bar\omega^k \bar\psi_{k}\bar\psi_{i}
+2\alpha\,x^{-1}\,\omega^k \bar\psi_{k} \psi_{i}
-x^{-1}\, \omega^k \bar\psi_i \psi_k \,,\label{4str-Psib}
\\ [4pt]
\label{4str-Z}
\delta \chi^{iA}&=&-2\alpha\,x^{-1}\left[\omega^{(i}\psi^{j)} +\bar\omega^{(i}\bar\psi^{j)}\right] \chi^{A}_j\,.
\end{eqnarray}
Once again, from \p{4str-L} and \p{4str-Z} we notice that the fermionic transformations of the fields  $\textsc{l}^{ia}$ and
$\chi^{iA}$ look as the field-dependent $SU(2)$ transformations of the doublet indices of these fields.

Proceeding from the component $D(2,1;\alpha)$ transformations \p{4str-X} - \p{4str-Z},
one can construct the corresponding Noether (super)charges
\begin{equation}\label{4Q-cl}
{Q}^i =p\, \psi^i- i x^{-1}\psi_{k}\,\Big[{\textstyle\frac{2}{3}}\,(1+2\alpha)\,\psi^{(i}\bar\psi^{k)}
- \alpha\,\chi^{iA} \chi^{k}_A
-2i\,\alpha\,\textsc{l}^{(i}{}_a p^{k)a} \Big]\, ,
\end{equation}
\begin{equation}\label{4Qb-cl}
\bar{Q}_i=p\, \bar\psi_i+i x^{-1}\bar\psi^{k}\,\Big[{\textstyle\frac{2}{3}}\,(1+2\alpha)\,\psi_{(i}\bar\psi_{k)}
- \alpha\,\chi_i^{A} \chi_{kA}
+ 2i\,\alpha\,\textsc{l}_{(i}{}^a p_{k)a} \Big]\,,
\end{equation}
\begin{equation}\label{4S-cl}
{S}^i =x \psi^i - t\,{Q}^i,\qquad \bar{S}_i=x\bar\psi_i-t\,\bar{Q}_i\,.
\end{equation}
Here $ p\equiv \dot x$ is the canonical momentum for $x$
and
\begin{equation}\label{mom-L}
p_{ia}=x^2\dot{\textsc{l}}_{ia}  -
i\alpha \,\textsc{l}^{k}{}_a\Big[ 2\psi_{(i}\bar\psi_{k)} - \alpha\chi_i^{A}\chi_{kA}\Big]
\end{equation}
are the momenta for $\textsc{l}^{ia}$. The momenta \p{mom-L} satisfy the constraint
\begin{equation}\label{costr-mom-L}
\textsc{l}^{ia}p_{ia}=0\,,
\end{equation}
which forms the pair of second class constraints together with the constraint \p{4N-440b},
\begin{equation}\label{costr-ell-L}
\textsc{l}^{ia}\textsc{l}_{ia}-{\textstyle\frac{1}{\alpha^2}}=0 \,.
\end{equation}

Like in subsection 5.2, the presence of second class constraints \p{costr-mom-L}, \p{costr-ell-L} implies that in the present case one
should use the Dirac brackets
\begin{equation}\label{DB-L}
\begin{array}{c}
\{ \textsc{l}^{ia}, \textsc{l}^{kb} \}_{{}_{DB}}= 0\,,\qquad
\{ \textsc{l}^{ia}, p_{\,kb} \}_{{}_{DB}}= \delta_{k}^{i}\delta_{b}^{a} -\alpha^2\textsc{l}^{ia}\textsc{l}_{\,kb}\,, \\ [7pt]
\{ p_{ia}, p_{\,kb} \}_{{}_{DB}}= \alpha^2\left( p_{ia} \textsc{l}_{\,kb} - p_{\,kb} \textsc{l}_{ia}\right).
\end{array}
\end{equation}
Making use of them, one can directly check  that the supercharges \p{4Qb-cl}, \p{4Q-cl}, \p{4S-cl} form $D(2,1;\alpha)$ algebra \p{DB-QQ}
with the Hamiltonian
\begin{eqnarray}
H &=&{\displaystyle{\textstyle\frac{1}{2}}\,p^2  + \frac{1}{x^{2}}\Bigg[\textstyle\frac{1}{2}}\,p^{ia}p_{ia}
+i\alpha \,\textsc{l}^{i}{}_{a} p^{\,ka}\Big( 2\psi_{(i}\bar\psi_{k)} - \alpha\chi_i^{A}\chi_{kA}\Big)
\nn [6pt]
&&\qquad\qquad\quad +\,{\textstyle\frac{1}{4}}\,(1+2\alpha) \,
\psi_{i}\psi^{i}\,\bar\psi_{k} \bar\psi^{k}
+\alpha  \,\psi_{i}\bar\psi_{k}\chi^{iA}\chi^{k}_A
-{\textstyle\frac{1}{4}}\,\alpha^2\,\chi_{i}^{A}\chi^{}_{kA}\chi^{iB}\chi^{k}_B \Bigg]\label{H-cl-440}
\end{eqnarray}
and the SU(2) generators
\begin{equation} \label{J-cl-440}
J^{ik} = -i\Big[ \psi^{(i}\bar\psi{}^{k)}-{\textstyle\frac12}\,\chi^{iA}\chi^k_{A}-i\,\textsc{l}^{i}{}_{a} p^{\,ka}\Big].
\end{equation}
The remaining even generators $K$, $D$ and $I^{i^\prime k^\prime}$ are defined by the same expressions as  in \p{K-cl}, \p{D-cl} and \p{I-cl}, \p{I-cl1}.

Note the relations
\begin{equation} \label{id-L}
\textsc{l}^{i}{}_a p^{\,ka}=\textsc{l}^{(i}{}_a p^{\,k)a}\,, \qquad
\textsc{l}^{i}{}_a p^{\,ka}\, \textsc{l}_{ib} p_k{}^{b}={\textstyle\frac{1}{2\alpha^2}}\,p^{\,ka}p_{\,ka} \,.
\end{equation}
With the help of them, one can find the explicit form of the Casimir operators \p{Cas} of the $su(1,1)$, $su(2)_R$ and $su(2)_L$ algebras for the
case under consideration
\begin{eqnarray}
T^2&=&
{\textstyle\frac{1}{8}}\,(1+2\alpha) \,
\psi_{i}\psi^{i}\,\bar\psi_{k} \bar\psi^{k}
+\textstyle\frac{1}{2}\,\alpha  \,\psi_{i}\bar\psi_{k}\left(\chi^{iA}\chi^{k}_A +2i\,\textsc{l}^{i}{}_{a} p^{\,ka}\right)
\nn [6pt]
&&
-{\textstyle\frac{1}{8}}\,\alpha^2\left(\chi_{i}^{A}\chi^{}_{kA}+2i\,\textsc{l}_{ia} p_{\,k}{}^{a}\right)
\left(\chi^{iB}\chi^{k}_B +2i\,\textsc{l}^{i}{}_{b} p^{\,kb}\right),  \label{Cas-11-440} \\ [6pt]
J^2&=&
{\textstyle\frac{3}{8}}\,
\psi_{i}\psi^{i}\,\bar\psi_{k} \bar\psi^{k}
+{\textstyle\frac{1}{2}}\,\psi_{i}\bar\psi_{k}\left(\chi^{iA}\chi^{k}_A +2i\,\textsc{l}^{i}{}_{a} p^{\,ka}\right) \nn [6pt]
&&
-{\textstyle\frac{1}{8}}\left(\chi_{i}^{A}\chi^{}_{kA}+2i\,\textsc{l}_{ia} p_{\,k}{}^{a}\right)
\left(\chi^{iB}\chi^{k}_B +2i\,\textsc{l}^{i}{}_{b} p^{\,kb}\right),
\label{Cas-2-440} \\ [6pt]
I^2&=&  -{\textstyle\frac{3}{8}}\,
\psi_{i}\psi^{i}\,\bar\psi_{k} \bar\psi^{k}
\,.\label{Cas-1-440}
\end{eqnarray}
Using these expressions and the expression
\begin{equation}\label{QQ-clCas-440}
{\textstyle\frac{i}{4}}\, Q^{ai^\prime i}Q_{ai^\prime i} ={\textstyle\frac{1}{2}}\,(1+2\alpha) \,
\psi_{i}\psi^{i}\,\bar\psi_{k} \bar\psi^{k}
+\alpha  \,\psi_{i}\bar\psi_{k}\left(\chi^{iA}\chi^{k}_A +2i\,\textsc{l}^{i}{}_{a} p^{\,ka}\right),
\end{equation}
we find that the second-order (classical) Casimir operator of $D(2,1;\alpha)\,$
takes the form
\begin{equation}\label{cl-Cas-12-440}
C_2=-{\textstyle\frac{1}{8}}\,\alpha(1+\alpha)\left(\chi_{i}^{A}\chi^{}_{kA}+2i\,\textsc{l}_{ia} p_{\,k}{}^{a}\right)
\left(\chi^{iB}\chi^{k}_B +2i\,\textsc{l}^{i}{}_{b} p^{\,kb}\right).
\end{equation}
Like in the previous cases, the additional fermionic variables $\chi_{i}^{A} $ make significant contributions
to the $D(2,1;\alpha)$, $su(1,1)$ and $su(2)_R$  Casimirs (\ref{Cas-11-440}), (\ref{Cas-2-440}), (\ref{cl-Cas-12-440}).

By the same tokens as in the previous cases, we obtain the various relations between the Casimirs pertinent
to the concrete realization of $D(2,1;\alpha)$ we have constructed in this section. These relations are
\begin{equation}\label{M-cl-440}
{M}\equiv {T}^2 - \alpha^2 {J}^2-  {\textstyle\frac{1}{3}}\,(1- \alpha^2)\,{I}^2
-{\textstyle\frac{i}{8}}\, (1-\alpha)\, {Q}^{ai^\prime i} {Q}_{ai^\prime i} =0\,,
\end{equation}
\begin{equation}\label{cl-constr-440}
(1+\alpha)\Big[T^2  - \alpha J^2 +{\textstyle\frac{1}{3}}\,(1-\alpha) I^2\Big]
+(1-\alpha) C_2=0 \,,
\end{equation}
\begin{equation}\label{H-constr-1-440}
T^2  = \alpha J^2 -{\textstyle\frac{1}{3}}\,(1-\alpha) I^2
+{\textstyle\frac{1}{8}}\,\alpha(1-\alpha)\left(\chi_{i}^{A}\chi^{}_{kA}+2i\,\textsc{l}_{ia} p_{\,k}{}^{a}\right)
\left(\chi^{iB}\chi^{k}_B +2i\,\textsc{l}^{i}{}_{b} p^{\,kb}\right).
\end{equation}
All these relations are valid for any $\alpha$, including $\alpha{=}-1\,$.

The Hamiltonian~(\ref{H-cl-440}) can be cast in the standard form of the Hamiltonian of (super)conformal mechanics
\begin{equation}\label{H-cl-st-440}
H ={\textstyle\frac{1}{2}}\,p^2  +
\frac{2T^2}{x^2} \,.
\end{equation}
Using the relation~(\ref{H-constr-1-440}), we can
rewrite it in the equivalent form as
\begin{eqnarray}
H &=&{\displaystyle
{\textstyle\frac{1}{2}}\,p^2   + \alpha \,
\frac{J^2}{x^2} - (1-\alpha) \, \frac{I^2}{3x^2}
} \nn [7pt]
&&{\displaystyle
+\,\alpha(1-\alpha)\,\frac{\left(\chi_{i}^{A}\chi^{}_{kA}+2i\,\textsc{l}_{ia} p_{\,k}{}^{a}\right)
\left(\chi^{iB}\chi^{k}_B +2i\,\textsc{l}^{i}{}_{b} p^{\,kb}\right)}{8x^2}\,.
}\label{H-cl-a-440}
\end{eqnarray}

\section{Superconformal coupling of the multiplets $({\bf 1, 4, 3})$ and $({\bf 4, 4, 0})$  mediated by
the  multiplet $({\bf 0, 4, 4})$}

Here we illustrate the efficiency of the off-shell superfield approach for constructing
new superconformal systems which involve dynamical supermultiplets of different types.
Here we  construct a system in which the superconformal interaction between the multiplets $({\bf 1, 4, 3})$ and $({\bf 4, 4, 0})$
arises as a result of coupling of these both multiplets to the single $({\bf 0, 4, 4})$ multiplet.

To this end,  we will consider a sum of the superfield actions (\ref{conf_v}), (\ref{Confpsi}) which describe the superconformal coupling
of the $({\bf 1, 4, 3})$ and $({\bf 0, 4, 4})$ multiplets and the actions (\ref{act-440}), (\ref{Confpsi-440}) which describe an analogous
coupling  of the $({\bf 4, 4, 0})$ and $({\bf 0, 4, 4})$ multiplets. After elimination of the auxiliary fields in this sum,
we will obtain a new superconformal coupling of the $({\bf 1, 4, 3})$ and $({\bf 4, 4, 0})$ multiplets mediated
by the multiplet $({\bf 0, 4, 4})$.

The corresponding off-shell component action is the sum of the component actions  (\ref{4N-X-WZ}), (\ref{coup-143-comp}), (\ref{4N-440-c}),
(\ref{vpsiComp2}). To distinguish between contributions of different actions, we substituted the coupling constant as $b \rightarrow b_1$
in (\ref{coup-143-comp}) and $b \rightarrow b_2$ in (\ref{vpsiComp2}).

Eliminating the auxiliary fields $A_{ik}$ and $F^A, \bar{F}_A$ from this sum by their equations of motion
\be
\begin{array}{rcl}
A_{ik}&=& i\left({\textstyle\frac{1}{\alpha}+2} \right)r^{-1}\varphi_{(i}\bar\varphi_{k)} -
4b_1\alpha^2i r^{\frac{1}{\alpha}+2}\phi_{(i}^{A}\phi_{k)A}\,,
\\ [6pt]
F^A&=& \Big[b_1r+b_2(f^2)^{-1}\Big]^{-1}\Big[b_1\varphi_{k}+2b_2(f^2)^{-2}f_{ka}\varphi^{a}\Big]\phi^{kA}\,,
\\ [6pt]
\bar{F}_A &=&\Big[b_1r+b_2(f^2)^{-1}\Big]^{-1}\Big[b_1\bar\varphi_{k}+2b_2(f^2)^{-2}f_{ka}\bar\varphi^{a}\Big]\phi^k_{A}\,,
\end{array}
\ee
we obtain the action with the Lagrangian
\begin{eqnarray}
L &=& \textstyle{\frac{1}{8\alpha^2}}\,
r^{-\frac{1}{\alpha}-2}\,\left[ \dot r\dot r +i \left(\bar\varphi_k \dot\varphi^k
-\dot{\bar\varphi}_k \varphi^k \right) \right]
+\textstyle{\frac{1}{2\alpha^2}}\,
(f^2)^{\frac{1}{\alpha}-1}\,\Big[ \dot f^{ia}\dot f_{ia} +\textstyle{\frac{i}{2}}\, \left(\bar\varphi_a \dot\varphi^a
-\dot{\bar\varphi}_a \varphi^a \right) \Big] \nonumber\\ [6pt]
&&
-i \left[b_1 r+b_2 (f^2)^{-1}\right]\phi^{iA}\dot{\phi}_{iA}  \nonumber \\ [6pt]
&& -\,\textstyle{\frac{i}{\alpha^2}}\,(\textstyle{\frac{1}{\alpha}}-1)\,
(f^2)^{\frac{1}{\alpha}-2}\,   f^{ia}\dot f^{b}_i \varphi_{(a}\bar\varphi_{b)}
-2i b_2 (f^2)^{-2} f_i^a\dot{f}_{ka} \,\phi^{iA}\phi_A^{k}
\nonumber \\ [6pt]
&&    + \,
\textstyle{\frac{1+2\alpha}{48\,\alpha^4}}\,
r^{-\frac{1}{\alpha}-4}\, \varphi_{(i}\bar\varphi_{k)} \varphi^{i}\bar\varphi^{k}
- \,\textstyle{\frac{1}{6\alpha^3}}\,
(\textstyle{\frac{1}{\alpha}}-1)\,
(f^2)^{\frac{1}{\alpha}-2}\, \varphi^{a}\bar\varphi^{b} \varphi_{(a}\bar\varphi_{b)}
\nonumber \\ [6pt]
&& -
\textstyle{\frac{b_1}{2\alpha}}\,r^{-1}\Big[b_1r+b_2(f^2)^{-1}\Big]^{-1}\Big[b_1r+b_2(1+2\alpha)(f^2)^{-1}\Big]\varphi_{i}\bar\varphi_{k}\phi^{iA}\phi^{k}_A
   \nonumber \\ [6pt]
&& -
4b_1b_2\,r\,(f^2)^{-3}\Big[b_1r+b_2(f^2)^{-1}\Big]^{-1}f_{ia}f_{kb}\varphi^{a}\bar\varphi^{b}\phi^{iA}\phi^{k}_A
   \nonumber \\ [6pt]
&& +
2b_1b_2\,(f^2)^{-2}\Big[b_1r+b_2(f^2)^{-1}\Big]^{-1}\Big(\varphi_{i}\bar\varphi^{a}-\bar\varphi_{i}\varphi^{a}\Big)f_{ka}\,\phi^{iA}\phi^{k}_A
   \nonumber \\ [6pt]
&&
+\alpha^2 (b_1)^2 r^{\frac{1}{\alpha}+2}\,\phi_{(i}^{A}\phi_{k)A}\phi^{iB}\phi^{k}_B \,.\label{1+4}
\end{eqnarray}

Redefining the variables as
\begin{equation}\label{1+4-x}
x_1 = r^{-\frac{1}{2\alpha}} \,, \qquad
x_2 = (f^2)^{\frac{1}{2\alpha}} \,,
\end{equation}
\begin{equation}\label{1+4-l}
\textsc{l}_{ia}={\textstyle\frac{1}{\alpha}}\,(f^2)^{-\frac{1}{2}}\,f_{ia} \,, \qquad  \textsc{l}^{ia}\textsc{l}_{ia}= {\textstyle\frac{1}{\alpha^2}} \,,
\end{equation}
\begin{equation}\label{1+4-p1}
\psi_1^{i} = -{\textstyle\frac{1}{2\alpha}}\,r^{-\frac{1}{2\alpha}-1} \varphi^{i}\,, \qquad
\bar\psi_{1i} = -{\textstyle\frac{1}{2\alpha}}\,r^{-\frac{1}{2\alpha}-1} \bar\varphi_{i}\,,
\end{equation}
\begin{equation}\label{1+4-p2}
\psi_2^{i} = {\textstyle\frac{1}{\alpha}}\,(f^2)^{\frac{1}{2\alpha}-1}\, f^{ia}\varphi_{a}\,, \qquad
\bar\psi_{2i} = -{\textstyle\frac{1}{\alpha}}\,(f^2)^{\frac{1}{2\alpha}-1}\, f_{ia}\bar\varphi^{a}\,,
\end{equation}
\begin{equation}\label{1+4-ch}
\chi^{iA} = \sqrt{2}\left[b_1 r+b_2 (f^2)^{-1}\right]^{\frac{1}{2}}\,\phi^{iA}\,,
\end{equation}
we bring the Lagrangian \p{1+4} into the form
\begin{eqnarray}
L &=& \textstyle{\frac{1}{2}}\,
\left[ \dot x_1\dot x_1+\dot x_2\dot x_2+ (x_2)^2\, \dot{\textsc{l}}^{ia}\dot{\textsc{l}}_{ia} +
i \left(\bar\psi_{1k} \dot\psi_1^k -\dot{\bar\psi}_{1k} \psi_1^k  +
\bar\psi_{2k} \dot\psi_2^k -\dot{\bar\psi}_{2k} \psi_2^k  \right)-i\chi^{iA}\dot{\chi}_{iA}\right]
\nonumber \\
&& +\,i\,\alpha \,
{\textsc{l}}_i^{\,a}\dot{\textsc{l}}_{ka} \left\{ 2\psi_2^{(i}\bar\psi_2^{k)} -
\alpha b_2\,(x_2)^{-2\alpha}\Big[b_1\,(x_1)^{-2\alpha}+b_2\,(x_2)^{-2\alpha} \Big]^{-1}\,\chi^{iA}\chi^{k}_A\right\}
\nonumber \\
&&    + \,
\textstyle{\frac{1}{3}}\,(1+2\alpha)\,(x_1)^{-2}\,
\psi_{1(i}\bar\psi_{1k)} \psi_1^{i}\bar\psi_1^{k}  \, - \,
\textstyle{\frac{2}{3}}\,(1-\alpha) \, (x_2)^{-2} \,\psi_{2(i}\bar\psi_{2k)} \psi_2^{i}\bar\psi_2^{k}
\nonumber \\
&&
-
\alpha b_1(x_1)^{-2\alpha-2}\Big[b_1(x_1)^{-2\alpha}+b_2(1+2\alpha)(x_2)^{-2\alpha} \Big]\Big[b_1(x_1)^{-2\alpha}+b_2(x_2)^{-2\alpha} \Big]^{-2}
\psi_{1i}\bar\psi_{1k}\chi^{iA}\chi^{k}_A
\nonumber \\
&&
-
2\alpha^2 b_1b_2\,(x_1)^{-2\alpha}(x_2)^{-2\alpha-2}\Big[b_1(x_1)^{-2\alpha}+b_2(x_2)^{-2\alpha} \Big]^{-2}
\psi_{2i}\bar\psi_{2k}\chi^{iA}\chi^{k}_A
\nonumber \\
&&
+
2\alpha^2 b_1b_2\,(x_1)^{-2\alpha-1}(x_2)^{-2\alpha-1}\Big[b_1(x_1)^{-2\alpha}+b_2(x_2)^{-2\alpha} \Big]^{-2}
\Big[\psi_{1i}\bar\psi_{2k}-\psi_{2i}\bar\psi_{1k}\Big]\chi^{iA}\chi^{k}_A
\nonumber \\
&&
+ \,\textstyle{\frac{1}{4}}\,\alpha^2 \,(b_1)^2\,
(x_1)^{-4\alpha-2} \Big[b_1(x_1)^{-2\alpha}+b_2(x_2)^{-2\alpha} \Big]^{-2} \chi_{(i}^{A}\chi_{k)A}\chi^{iB}\chi^{k}_B \, . \label{total-1+4}
\end{eqnarray}

The on-shell supersymmetry transformations leaving \p{total-1+4} invariant up to a total derivative
are as follows
\begin{equation}\label{4str-X-c}
\delta x_1=-\omega_i\psi_1^i+ \bar\omega^i\bar\psi_{1i}\,, \qquad
\delta x_2=-\omega_i\psi_2^i+ \bar\omega^i\bar\psi_{2i}\,,
\end{equation}
\begin{equation}\label{4str-L-c}
\delta \textsc{l}^{ia}=-2\alpha\,(x_2)^{-1}\left[\omega^{(i}\psi_2^{j)} +\bar\omega^{(i}\bar\psi_2^{j)}\right]\textsc{l}_j{}^a
\,,
\end{equation}
\begin{eqnarray}
\delta \psi_1^i&=&i \bar\omega^i\dot x_1 + i\bar\eta^i x_1
-(1+2\alpha)\,(x_1)^{-1}\left(\omega_k \psi_1^{k}\psi_1^{i}+\bar\omega_k \psi_1^{k} \bar\psi_1^{i} \right) \nn
&&
+\,\alpha\,b_1\,(x_1)^{-2\alpha-1}\Big[b_1\,(x_1)^{-2\alpha}+b_2\,(x_2)^{-2\alpha} \Big]^{-1}\, \bar\omega_k \chi^{A(i} \chi^{k)}_A \,,
\nn
\delta \bar\psi_{1i}&=&-i \omega_i\dot x_1-i\eta_i x_1
+(1+2\alpha)\,(x_1)^{-1}\left(\bar\omega^k \bar\psi_{1k}\bar\psi_{1i}+\omega^k \bar\psi_{1k} \psi_{1i} \right) \nn
&&
+\alpha\,b_1\,(x_1)^{-2\alpha-1}\Big[b_1\,(x_1)^{-2\alpha}+b_2\,(x_2)^{-2\alpha} \Big]^{-1}\, \omega^k \chi^{A}_{(i} \chi_{k)A} \,,\label{2str-Psi}
\end{eqnarray}
\begin{eqnarray}
\delta \psi_2^i&=&i \bar\omega^i\dot x_2 + i\bar\eta^i x_2 -2\alpha i\, x_2\,\textsc{l}^{ia}\dot{\textsc{l}}^{k}{}_{a}\,\bar\omega_k \nn
&& -(1+2\alpha)\,(x_2)^{-1}\,\omega_k \psi_2^{k}\psi_2^{i}
-2\alpha\,(x_2)^{-1}\,\bar\omega_k \psi_2^{k} \bar\psi_2^{i}
+(x_2)^{-1}\, \bar\omega_k \psi_2^i \bar\psi_2^k   \,,
\nn
\delta \bar\psi_{2i}&=&-i \omega_i\dot x_2-i\eta_i x_2 -2\alpha i\, x_2\,\textsc{l}_{i}{}^{a}\dot{\textsc{l}}_{ka}\,\omega^k \nn
&&  +\,(1+2\alpha)\,(x_2)^{-1}\,\bar\omega^k \bar\psi_{2k}\bar\psi_{2i}
+2\alpha\,(x_2)^{-1}\,\omega^k \bar\psi_{2k} \psi_{2i}
-(x_2)^{-1}\, \omega^k \bar\psi_{2i} \psi_{2k} \,,\label{4str-Psi-c2}
\end{eqnarray}
\begin{eqnarray}
\delta \chi^{iA}&=&-2\alpha\,\Big[b_1\,(x_1)^{-2\alpha}+b_2\,(x_2)^{-2\alpha} \Big]^{-1}
\Bigg\{\omega^{(i}\left[b_1(x_1)^{-2\alpha-1}\psi_1^{j)}+b_2(x_2)^{-2\alpha-1}\psi_2^{j)}\right]
\nn
&&
\qquad\qquad\qquad\qquad\qquad\quad
+\bar\omega^{(i}\left[b_1(x_1)^{-2\alpha-1}\bar\psi_1^{j)}+b_2(x_2)^{-2\alpha-1}\bar\psi_2^{j)}\right] \Bigg\} \chi^{A}_j\,.\label{4str-Z-c}
\end{eqnarray}
It is straightforward to  compute the relevant   $D(2,1;\alpha)$ Noether (super)charges and Casimirs,
as it was done in the previous sections for the separate pair couplings. We leave finding the explicit expressions for the future.

Let us dwell on some peculiarities of the system constructed.

First,  the component action  associated with \p{total-1+4} involves a non-trivial dependence on both coupling constants $b_1$ and $b_2$,
in contrast to the actions for the separate pairs which are reproduced by setting $b_1 = 0$ or $b_2=0$.

Accordingly, we observe that the quartic fermionic terms in \p{total-1+4} have a more complicated dependence on $x_1$ and $x_2$
as compared to the separate $({\bf 1, 4, 3}) \oplus ({\bf 0, 4, 4})$ or $({\bf 4, 4, 0}) \oplus ({\bf 0, 4, 4})$ couplings,
in which cases we are left with the standard conformal denominators $(x_1)^{-2}$ or $(x_2)^{-2}$.

These peculiarities survive in the special cases $\alpha = -1/2$ corresponding to the superconformal group $OSp(2|4)$
and $\alpha = -1$ which corresponds to $PSU(1,1|2)$ and was the subject of study in \cite{Galaj}. It would be interesting
to reveal implications of this and, perhaps, some other ``hybrid'' $D(2,1;\alpha)$ invariant $d=1$ systems in the AdS$/$CFT and
supersymmetric black  hole business along the lines of Ref. \cite{Galaj} and related references.

\section{Concluding remarks}
In this paper we have presented in detail the construction of the realizations of the most general ${\cal N}=4, d=1$ superconformal
symmetry $D(2,1;\alpha)$ in the SCM models associated with the reducible ${\cal N}=4$ multiplets  $({\bf 1, 4, 3}) \oplus ({\bf 0, 4, 4})$,
$({\bf 3, 4, 1}) \oplus ({\bf 0, 4, 4})$   and   $({\bf 4, 4, 0}) \oplus ({\bf 0, 4, 4})$. In all cases, we started
from the manifestly supersymmetric off-shell superfield description, then passed to the off-shell component actions and,
finally, to the on-shell actions by eliminating the relevant sets of the auxiliary fields. Though the superfield  description
for the separate multiplets entering the various pairs was known before, neither full off- and on-shell superconformal
Lagrangians nor the explicit realizations of the $D(2,1;\alpha)$ generators for them at arbitrary $\alpha $ were given.
We also worked out an instructive example of the $D(2,1;\alpha)$ invariant action involving the multiplets $({\bf 1, 4, 3})$ and $({\bf 4, 4, 0})$
interacting through couplings to the multiplet $({\bf 0, 4, 4})$. All the models constructed admit a simple extension to an arbitrary number
of the multiplets $({\bf 0, 4, 4})$, like in the $\alpha =-1$ case treated in \cite{Galaj}.

The common feature of all models considered is the splitting of the involved variables into the radial (``dilaton'') part presented
by the field $x$ and the angular part containing everything else, including the fermionic variables of the multiplets $({\bf 0, 4, 4})$.
The structure of supercharges and Hamiltonians in terms of this splitting is basically universal  for all considered systems, like in their
$\alpha = -1$ particular case, and is, presumably, in the agreement with the general structure of ${\cal N}=4$ SCM models with $D(2,1;\alpha)$
invariance suggested in \cite{hkln}. As distinct from \cite{hkln}, we begin with the well defined reducible
off-shell ${\cal N}=4, d=1$ representation and come to the final expressions for the $D(2,1;\alpha)$ generators and
the splitting of variables just mentioned through the standard Noether procedure applied to the relevant invariant Lagrangians
and by making the universal field redefinition in the end. Another difference is that our off-shell approach, as is illustrated in section 7,  allows
one to gain, as a result of elimination of the auxiliary fields,  some additional fermionic couplings which are difficult to guess
within the intrinsically on-shell Hamiltonian approach.

It is of interest to apply the same methods to construct more general superconformal ${\cal N}=4$ models, involving, e.g., some mirror (or ``twisted'')
analogs \cite{1}, \cite{2}  of the multiplets considered here and to generalize our study to the $D(2,1;\alpha)$ invariant models with the ``trigonometric'' realization of
the conformal $d=1$ symmetry along the lines of Ref. \cite{ISTconf}. It is also desirable to consider quantum versions of all these models and $D(2,1;\alpha)$ realizations,
as well as to understand in full generality their possible links with the higher-dimensional supergravity, black holes and AdS/CFT in the spirit of refs.
\cite{nscm}, \cite{GTow}, \cite{Galaj}, \cite{TO}  and other related works.

Finally, we focus on the two surprising common features of the models presented.

One unusual property is seen already at the level when the extra $({\bf 0, 4, 4})$ multiplets are suppressed.
In the three cases considered in sections 4, 5 and 6 the corresponding Lagrangians are (without WZ terms for the $({\bf 3, 4, 1})$ multiplet)
\begin{eqnarray}
L_{(1,4,3)}&=&\textstyle{\frac{1}{2}}\left[ \dot x\dot x +i \left(\bar\psi_k \dot\psi^k
-\dot{\bar\psi}_k \psi^k  \right) \right]  + \textstyle{\frac{1}{3}}\,(1+2\alpha)\,x^{-2}\,
\psi^{i}\bar\psi^{k}  \psi_{(i}\bar\psi_{k)}\,,
\\ [7pt]
L_{(3,4,1)}&=&\textstyle{\frac{1}{2}}\left[ \dot x\dot x+  x^2 \dot \ell^{ik}\dot \ell_{ik}  +i \left(\bar\psi_k \dot\psi^k
-\dot{\bar\psi}_k \psi^k  \right) \right]+4i\alpha \, \ell_i^{\,j}\dot{\ell}_{kj} \psi^{(i}\bar\psi^{k)} \nn [5pt]
&&
-\, \textstyle{\frac{1}{3}}\,(1-2\alpha)x^{-2}
\psi^{i}\bar\psi^{k} \psi_{(i}\bar\psi_{k)} \,,
\\[7pt]
L_{(4,4,0)}&=&\textstyle{\frac{1}{2}}\left[ \dot x\dot x+  x^2 \dot {\textsc{l}}^{ia}\dot {\textsc{l}}_{ia}  +i \left(\bar\psi_k \dot\psi^k
-\dot{\bar\psi}_k \psi^k  \right) \right]+2i\alpha \, {\textsc{l}}_i^{\,a}\dot{\textsc{l}}_{ka} \psi^{(i}\bar\psi^{k)} \nn [5pt]
&&
-\, \textstyle{\frac{2}{3}}\,(1-\alpha)x^{-2}
\psi^{i}\bar\psi^{k} \psi_{(i}\bar\psi_{k)} \,,
\end{eqnarray}
where $x$ is the radial variable and $\ell_{ik}\,$, ${\textsc{l}}_{ia}$ are angular variables. We see,
that the four-fermionic terms are vanishing at {\it different} values of $\alpha$ for different multiplets:
$\alpha=-1/2$ for the multiplet $({\bf 1, 4, 3})$, $\alpha=1/2$ for the multiplet $({\bf 3, 4, 1})$ and $\alpha=1$ for the multiplet $({\bf 4, 4, 0})$.

On the other hand, in the supercharges
\begin{eqnarray}
{Q}^i_{(1,4,3)}&=&p\, \psi^i- {\textstyle\frac{2i}{3}}\,(1+2\alpha)\,x^{-1}\psi_{k}\,\psi^{(i}\bar\psi^{k)}\,,
\\ [7pt]
{Q}^i_{(3,4,1)}&=&p\, \psi^i- {\textstyle\frac{2i}{3}}\,(1+2\alpha)\,x^{-1}\psi_{k}\,\psi^{(i}\bar\psi^{k)}
-4\alpha\,x^{-1}\psi_{k}\,\ell^{(i}_j p^{\,k)j} \,,
\\[7pt]
{Q}^i_{(4,4,0)}&=&p\, \psi^i- {\textstyle\frac{2i}{3}}\,(1+2\alpha)\,x^{-1}\psi_{k}\,\psi^{(i}\bar\psi^{k)}
-2\alpha\,x^{-1}\psi_{k}\,\textsc{l}^{(i}{}_a p^{\,k)a}
\end{eqnarray}
the three-fermion terms vanish at $\alpha=-1/2$ for {\it all} multiplets we dealt with. Moreover, the angular {\it dynamical} variables
in the supercharges can be absorbed into the relevant SU(2) currents.

The second common feature is that the contribution of the $({\bf 0, 4, 4})$ multiplets to the supercharges is universal for all multiplets:
\begin{eqnarray}
{Q}^i_{(1,4,3)\oplus(0,4,4)}&=&p\, \psi^i- i x^{-1}\psi_{k}\,\Big[{\textstyle\frac{2}{3}}\,(1+2\alpha)\,\psi^{(i}\bar\psi^{k)}
- \alpha\,\chi^{iA} \chi^{k}_A \Big]\,,
\\ [7pt]
{Q}^i_{(3,4,1)\oplus(0,4,4)}&=&p\, \psi^i- i x^{-1}\psi_{k}\,\Big[{\textstyle\frac{2}{3}}\,(1+2\alpha)\,\psi^{(i}\bar\psi^{k)}
-4i\,\alpha\,\ell^{(i}_j p^{\,k)j} - \alpha\,\chi^{iA} \chi^{k}_A \Big] \,,
\\[7pt]
{Q}^i_{(4,4,0)\oplus(0,4,4)}&=&p\, \psi^i- i x^{-1}\psi_{k}\,\Big[{\textstyle\frac{2}{3}}\,(1+2\alpha)\,\psi^{(i}\bar\psi^{k)}
-2i\,\alpha\,\textsc{l}^{(i}{}_a p^{\,k)a} - \alpha\,\chi^{iA} \chi^{k}_A \Big]\,.
\end{eqnarray}

It would be interesting to learn whether these properties survive quantization.

\section*{Acknowledgements}
We acknowledge support from the RFBR grant 15-02-06670 and a grant of the Heisenberg - Landau program.
E.I. thanks Anton Galajinsky for a correspondence which has revived his interest in this circle of problems.

\end{document}